\documentclass[11pt]{rspublic}
\def\HL#1{#1}
\def\DV#1{#1}

\usepackage{natbib}

\def\cite{\citep}
\usepackage{feynmf}%%%{feynmp}

\usepackage{geometry}                % See geometry.pdf to learn the layout options. There are lots.
\usepackage{graphicx}
\usepackage{amssymb}
\usepackage{epstopdf}
\usepackage{subfigure}
\usepackage{wrapfig}
\newcommand\field[1]{{\ensuremath{\mathbb{{#1}}}}}

\usepackage{tikz}
\usetikzlibrary{arrows}
\tikzstyle{block}=[draw opacity=0.7,line width=1.4cm]
\tikzstyle{theorem}=[draw opacity=0.7,line width=1.4cm]

\newcommand{\vev}[1]{\langle#1 \rangle}
\definecolor{darkgreen}{rgb}{0,0.4,0}
\definecolor{darkred}{rgb}{0.4,0,0}
\definecolor{darkblue}{rgb}{0,0,0.4}

\newcommand{\greencom}[1]{\textcolor{darkgreen}{\footnotesize{#1}}}

\def\bl{\textcolor{darkblue}}
\def\redt{\textcolor{darkred}}
\def\greent{\textcolor{darkgreen}}

\def\bpsi{\bar\psi}

\def\p{\partial}

\def\half{{1\over 2}}

\def\Im{{\rm Im\hskip0.1em}}

\def\vev#1{\langle{#1}\rangle}
\def\Dslash{\rlap{\hskip0.2em/}D}

\def\CC{{\cal C}}

\def\CO{{\cal O}}%AEL

\def\II{\relax{I\kern-.10em I}}

\def\IZ{\relax{\rm Z\kern-.34em Z}}
\def\IB{\relax{\rm I\kern-.18em B}}
\def\IC{{\relax\hbox{$\inbar\kern-.3em{\rm C}$}}}
\def\ID{\relax{\rm I\kern-.18em D}}
\def\IE{\relax{\rm I\kern-.18em E}}
\def\IF{\relax{\rm I\kern-.18em F}}
\def\IG{\relax\hbox{$\inbar\kern-.3em{\rm G}$}}
\def\IGa{\relax\hbox{${\rm I}\kern-.18em\Gamma$}}
\def\IH{\relax{\rm I\kern-.18em H}}
\def\II{\relax{\rm I\kern-.18em I}}
\def\IK{\relax{\rm I\kern-.18em K}}
\def\IP{\relax{\rm I\kern-.18em P}}
\def\inbar{\,\vrule height1.5ex width.4pt depth0pt}

\def\p{\partial}

\def\IR{\relax{\rm I\kern-.18em R}}

\def\lp10{\ell_p^{10}}
\def\lp11{\ell_p^{11}}
\def\R11{R_{11}}

\def\frac#1#2{{#1 \over #2}}

\def\({\left(}
\def\){\right)}
\def\[{\left[}
\def\]{\right]}

\def\eg{{\it e.g.}}
\def\ie{{\it i.e.}}

\def\sG{{\cal G}}
\def\om{\omega}
\def\ov{\over}

\newcommand{\vk}{{\vec k}}

\newcommand{\RR}{\field{R}}

\newcommand\sig{\sigma}

\newcommand\Lam{\Lambda}

\newcommand\Ga{{\ensuremath{{\Gamma}}}}

\newcommand\De{{\ensuremath{{\Delta}}}}

\def\yandz{\Phi}
\def\ri){\right)}
\def\le({\left(}
\def\tildem{\tilde m}
\newcommand{\bca}{\begin{cases}}
\newcommand{\eca}{\end{cases}}

\newcommand{\be}{\begin{equation}}
\newcommand{\ee}{\end{equation}}

\newcommand{\bwt}{\begin{widetext}}
\newcommand{\ewt}{\end{widetext}}

\newcommand{\ben}{\begin{enumerate}}
\newcommand{\een}{\end{enumerate}}

\newcommand{\bln}{\begin{align}}
\newcommand{\eln}{\end{align}}
\newcommand{\bst}{\begin{split}}
\newcommand{\est}{\end{split}}

\newcommand{\bea}{\begin{eqnarray}}
\newcommand{\eea}{\end{eqnarray}}

\newcommand\sO{{\ensuremath{{\mathcal O}}}}

\newcommand{\largeN}{\textcolor{darkred}{N^2}}

\def\spinor{\psi}
\def\scalar{\varphi}
\def\mspinor{m_{\spinor}}
\def\qspinor{q_{\spinor}}
\def\mscalar{m_{\scalar}}
\def\qscalar{q_{\scalar}}

\title{Holographic non-Fermi liquid fixed points}
\author[Tom Faulkner,
Nabil Iqbal, Hong Liu, John McGreevy and David Vegh]{Tom Faulkner$^{1}$,
Nabil Iqbal$^2$, Hong Liu$^{2}$, John McGreevy$^{2}$ and David Vegh$^{3}$}
\affiliation{
${}^1${KITP, Santa Barbara, CA 93106},\\
${}^2${Center for Theoretical Physics,
Massachusetts Institute of Technology,
Cambridge, MA 02139},
${}^3${Simons Center for Geometry and Physics, Stony Brook University, Stony Brook, NY 11794-3636}}
\begin{document}
\maketitle
%\section{}
%\subsection{}

\label{firstpage}
\begin{abstract}%{KEYWORDS GO HERE}

Techniques arising from string theory
can be used to study assemblies of strongly-interacting fermions.
Via this `holographic duality',
various strongly-coupled many body systems
are solved using an auxiliary theory of gravity.
Simple holographic realizations of finite density
exhibit single-particle spectral functions with sharp Fermi surfaces,
of a form distinct from those of the Landau theory.
The self-energy is given by a correlation function in an infrared fixed point theory which is represented by an 
$AdS_2$ region in the dual gravitational description. Here we describe in detail the gravity calculation of this IR correlation function.

This article is a contribution to a special issue of {\it Phil.~Trans.~A} on the normal state of the cuprates;
as such, we also provide some review and context.
\end{abstract}

\section{Introduction}

The metallic
%{\it{metallic}}
states
%of a finite density of fermions
that we understand well are described by Landau Fermi liquid theory.
This is
%\greencom{(It describes normal metals, ${\rm He}_3$, ...)}
%\hskip0.1in\includegraphics[scale=0.09]{pictures/fermi_sea.eps}
a free stable RG fixed point
\citep{Benfatto:1990zz, Polchinski:1992ed, Shankar:1993pf}
(modulo the BCS instability which sets in at parametrically low
temperatures).
Landau quasiparticles manifest themselves as (a surface of)
poles in single-fermion Green's function
at $k_\perp  \equiv |\vec k| -  k_F = 0 $
%$$ G_R (t, \vec x) =
%i \theta(t) \vev{ \{  \psi^\dagger(t, \vec x)  , \psi(0, \vec 0) \} }_\mu $$
\be G_R(\omega, k) = { Z \over   \om - v_F k_\perp  + i \Gamma} + ...
%~~~ k_\perp \equiv |\vec k| - k_F 	
\ee
where the dots represent incoherent contributions.
Landau quasiparticles are long-lived: their width is
$\Gamma \sim {\omega_\star^2} $,
where $\omega_\star(k)$ is the
real part of the location of the pole.
%frequency of the excitation.
%Because of the kinematics of the Fermi surface, they only interact by irrelevant operators.
The residue $Z$, their overlap with the external electron,
is finite on Fermi surface,
and the spectral density becomes arbitrarily sharp there
%The spectral density encodes
\be
A(\omega, k ) \equiv { 1\over \pi} \Im G_R(\omega, k)
\buildrel{k_\perp \to 0}\over \to
Z \delta(\omega - v_F k_\perp) ~~.\ee
 Thermodynamical and  transport behavior of the system can be characterized in terms of these
long-lived quasi-particles.
%Reliable calculation of thermodynamics and  transport relies on the
%presence of this delta function.

{Non-Fermi liquid metals (NFL) exist but are mysterious};
the `strange metal' phase of optimally-doped cuprates
is a notorious example.
%\begin{center}
%\includegraphics[height=70pt]{pictures/high_tc1.eps}
%\hskip0.3in
%\includegraphics[height=50pt]{pictures/ARPES1.eps}
%\hskip-0.3in
%$ \Longrightarrow$
%\includegraphics[scale=0.3]{pictures/ZXShen.eps}
%\end{center}
%\cite{ZX Shen}
For example, data from angle-resolved photoemission (ARPES)
in this phase 
(see \citep{damascelli} and references therein)
indicate gapless modes -- the spectral function $A (\om, k)$ exhibits
nonanalyticity at $\omega \sim 0, k \sim k_F$ -- around some $k_F$ with
 width $\Gamma(\omega_\star) \sim \omega_\star$,
%vanishing residue $Z \buildrel{\omega \to 0} \over{\sim}{1\over |\ln \omega|}.$
and vanishing residue $Z \buildrel{k_\perp \to 0} \over{\to}0.$
Furthermore, the system's resistivity exhibits a linear temperature dependence in sharp contrast to the quadratic dependence of a Fermi liquid. These anomalies , along with others, suggest that in the strange metal phase,
there is still a sharp Fermi surface, but no long-lived quasiparticles.

Known field theoretical examples of non-Fermi liquids which also exhibit the phenomenon of a Fermi surface without long-lived quasi-particles include Luttinger liquids in 1+1 dimensions \cite{affleck}, and a free fermion gas coupled to some gapless bosonic excitations, which can be either a transverse gauge field or certain order parameter fluctuations near a quantum critical point~(see \cite{Holstein:1973zz,reizer,baym,Polchinski:1993ii,Nayak:1993uh,Halperin:1992mh,altshuler:1994,Schafer:2004zf,
Boyanovsky,nagaosa,Patrick,fradkin,lawler,nave,sungsikgaugefield,Metlitski:2010pd,Mross:2010rd}).
Neither of these, however, is able to explain the behavior of a strange metal phase. The former is specific to $(1+1)$-d kinematics. In the latter class of examples, the influence of gapless bosons is mostly along the forward direction, and is not enough, for example, to account for the linear temperature dependence of the resistivity.

\bigskip

Here we summarize recent findings of a class of non-Fermi liquids, some of which share similar low energy behavior to those of a strange metal phase, using the holographic approach~\cite{Liu:2009dm, Faulkner:2009wj, Faulkner:2009am,Faulkner:2010da, resistivitypaper}
(see also~\cite{Lee:2008xf, Cubrovic:2009ye, soojong}). One of the most intriguing aspects of these systems is that their low energy behavior is controlled by an infrared (IR) fixed point, which exhibits nonanalytical behavior only in the time direction. In particular, single-particle spectral function and charge transport can be characterized by the scaling dimension of the fermionic operator in the IR fixed point.

We should emphasize that at a microscopic level the systems we consider differ very much from the electronic systems underlying strange metals:  they are translationally invariant and spherically symmetric;
they typically involve a large number of fermions, scalars and gauge fields, characterized by a parameter $N$, and we have to work with the large $N$ limit;  at short distances, the systems approach a relativistic conformal field theory (CFT). Nevertheless, the similarity of their low energy behavior to that of a strange metal is striking and may not be an accident. After all, the key to our ability to characterize many-body systems has often been
universalities of low energy physics among systems with different microscopics.

We should also mention that the holographic non-Fermi liquids we describe here
likely reflect some intermediate-scale physics rather than genuine ground states.
The systems in which they are embedded can have various superconducting~\cite{holographicsc1,holographicsc2}\footnote{For a review see \cite{holographicsc3} and \cite{holographicsc4}.} or magnetic~\cite{Iqbal:2010eh} instabilities, just as in real-life condensed matter systems.

In the next section we give a lightning introduction to holographic duality.
The subsequent two sections describe the structure of
the fermionic response in the simplest holographic realization of finite density.
\S\ref{sec:cor} is the core of this paper, where we describe
in detail the calculation of the self-energy.
In the final sections we comment on transport, the superconducting state,
and a useful cartoon of the mechanism which kills the quasiparticles.

\section{Holographic duality}

%String theory has produced something valuable.

The study of black holes has taught us
that a quantum theory of gravity has a number
of degrees of freedom that is sub-extensive.
In general, one expects a gravitational system in some volume
to be describable
in terms of an ordinary quantum system living on the boundary.
For a certain class of asymptotic geometries,
these boundary degrees of freedom has been
precisely identified \cite{AdS/CFT1,AdS/CFT2,AdS/CFT3}.
For a review of this
`holographic duality'
in the present spirit, see
\cite{Sachdev:2008ba, Hartnoll:2009sz, McGreevy:2009xe, Hartnoll:2009qx, Sachdev:2010ch}.

The basic example of the duality is the following.
A theory of gravity in $d+1$-dimensional Anti-de Sitter space, $AdS_{d+1}$,
{\it is} a
conformal field theory in $d$ spacetime dimensions.
Here by `is' we mean that the observables are in one-to-one correspondence.
%{(many generalizations, CFT is best-understood.)}
$AdS$ space,
\be
\label{eq:ads}
ds^2 = {r^2 \over R^2} \( \textcolor{darkgreen}{-dt ^2 + d\vec x^2} \) + R^2 { dr^2 \over r^2} ~,\ee
can be viewed a collection of copies of
Minkowski space $\RR^{d,1}$ (whose isometries are the Poincar\'e group)
of varying `size', parametrized by a coordinate $r$.
The isometries of $AdS_{d+1}$ are precisely the conformal group in $d$ spacetime dimensions.
The `boundary' where the dual field theory lives is at $r = \infty$ in these coordinates.

\begin{figure}
\begin{center}
\hskip-.1in
\includegraphics[height=100pt]{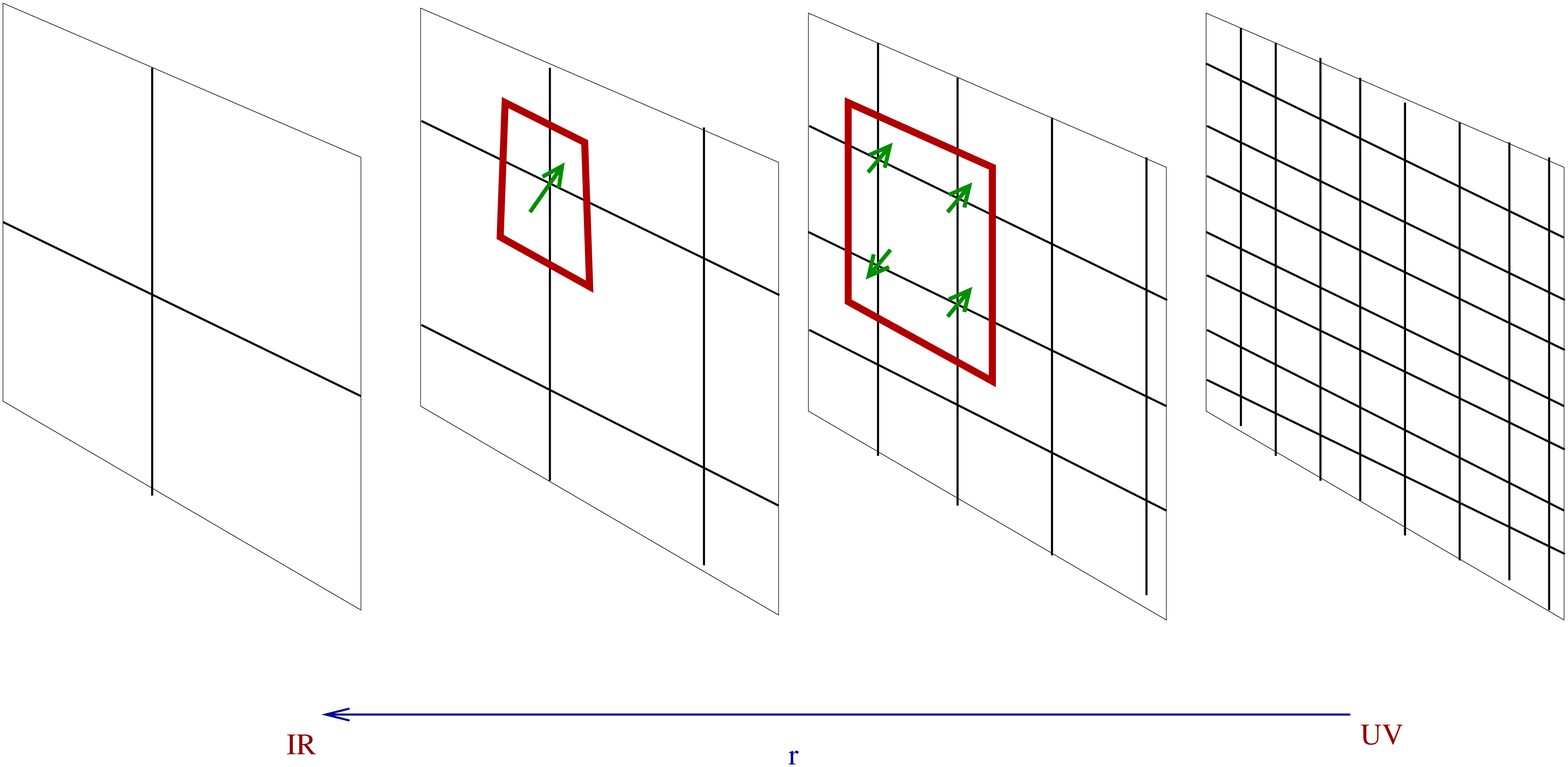} % requires the graphicx package
%$\longrightarrow$
\hskip0.3in
\includegraphics[height=120pt]{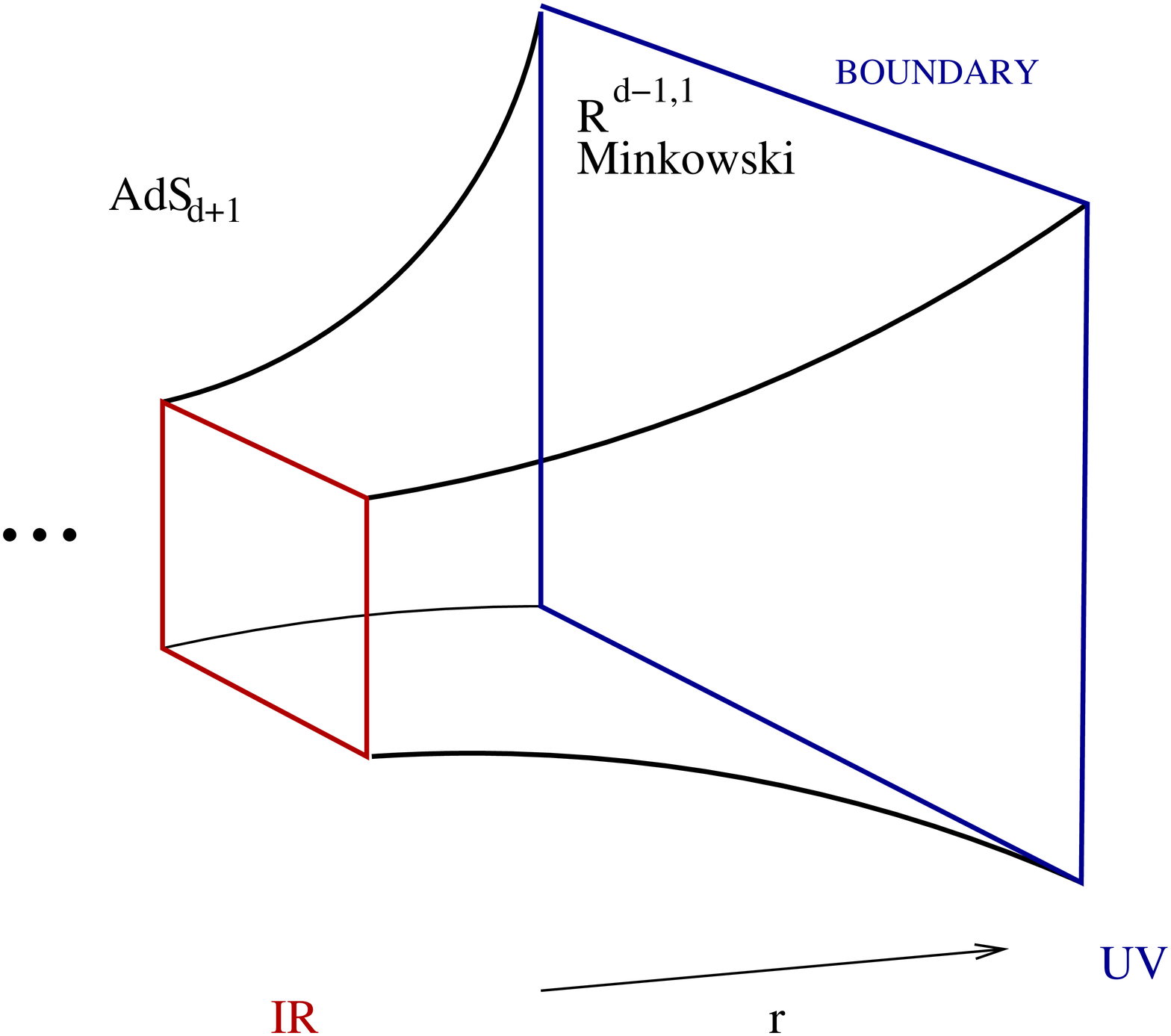} % requires the graphicx package
\caption{
The left figure indicates a series of block spin transformations labelled by a parameter $r$.
The right figure is a cartoon of AdS space, which organizes the field theory information
in the same way.
In this sense, the bulk picture is a hologram:
excitations with different wavelengths get put in different places in the bulk image.
}
\end{center}
\end{figure}
The extra (`radial') dimension can be considered as corresponding to  the resolution scale for the dual theory. The evolution of the geometry (and the fields propagating therein)
along this radial direction
represent the RG flow of the dual field theory.
The `AdS radius', $R$ in \eqref{eq:ads},
is a dimensionful parameter which, roughly, encodes the rate of RG flow. 
%\footnote{\HL{This sentence should be removed. not sure it makes sense. }}

%When is this correspondence useful?
A practical consequence of this duality is the following equation
for the field theory partition function:
\bea
\label{GKPW}
 Z_{QFT}[{\rm sources}] &=& Z_{{\rm quantum~gravity}}[{\tiny{\rm b.~c.~at~}r\to \infty}]
\\  \nonumber
&\approx& e^{ - \textcolor{darkred}{N^2} S_{{\rm bulk}}[{\rm b.~c.~at~}r\to \infty]}|_{{\rm extremum~of~}S_{{\rm bulk}}}
% \textbf{Inner:} \quad  \sigma \equiv \omega {R_2^2 \ov r-1} \quad {\rm for} \quad
% \epsilon < \sigma < \infty \\
% \nonumber
%&& \textbf{Outer:} \quad  {\omega R_2^2 \ov \epsilon}  < r-1
\eea
The RHS of the first line is a partition function of quantum gravity,
a general understanding of which is still lacking. \DV{The boundary conditions at $r\to \infty$ specify the sources in the QFT partition function. }
In the second line, \HL{the classical approximation to  the quantum gravity partition function
has been made via a saddle point approximation. The dimensionless parameter $\textcolor{darkred}{N^2}$, which makes such an approximation possible, is the inverse of the Newton constant in units of curvature radius $R$, and is large in the classical limit. Through the duality, $N^2$ is mapped to the number of degrees of freedom per point in the dual free field theory.}
%The description in terms of classical gravity
%is valid
%(the saddle point is sharply defined)
%when the field theory has
%many degrees of freedom per point, $\textcolor{darkred}{N^2} \gg 1$.
In the best-understood examples,
the classical nature of the gravity theory
arises from large-$N$ factorization
%the limit where the gravity theory becomes classical
in the 't Hooft limit of a gauge theory.

Fields in $AdS_{d+1}$ correspond to operators in CFT;
the mass of the field determines the scaling dimension of the operator.
The boundary conditions on a bulk field specifies the
coefficient of the corresponding operator in the field theory action.
For example, since the bulk theory
is a gravitational theory, the bulk spacetime metric $ds^2 = g_{\mu\nu}dx^\mu dx^\nu$ is dynamical.
The boundary value of bulk metric $\lim_{r \to \infty} g_{\mu\nu} $
is the source for the field theory stress-energy tensor
$T^{\mu\nu}$.
So we can rewrite \eqref{GKPW} as
\be
\vev{e^{ \phi_a^{(0)} \CO^a }  }
\approx e^{ - \textcolor{darkred}{N^2} S_{{\rm bulk}}[ \phi_a
\buildrel{r\to\infty}\over{\to} \phi_a^{(0)} ] }|_{{\rm extremum~of~}S_{{\rm bulk}}}~~.
\ee
We can change the field theory action just by changing
the {\it boundary conditions} on the bulk fields.
Different couplings in the bulk action then correspond to entirely different field theories.
For example, changing the bulk Newton constant (\HL{in units of curvature radius})
is accomplished by changing the parameter $\largeN$;
but this is the number of degrees of freedom per point of the field theory.

In addition to large $\largeN$, useful calculation requires the background geometry
to have small curvature.
This is because otherwise we cannot reliably approximate the
bulk action as
\be\label{bulkactionwithoutgaugefield}
S = \int d^{d+1} x \sqrt g ~ \( {\cal R} + {d (d+1) \over R^2} + ... \)
\ee
where ${\cal R}$ is the Ricci scalar, and
the ellipsis indicates terms with more powers of the curvature
(and hence more derivatives).
One necessary condition for this is that the
`$AdS$ radius' $R$ be large compared to the
energy scale set by the string tension.
Given how it appears in \eqref{bulkactionwithoutgaugefield},
this requirement is a bulk version of the cosmological constant problem --
the vacua with large cosmological constant require more information about
the full string theory.
The largeness of the dual geometry implies that the dual QFT is
strongly coupled.
Circumstantial evidence for this statement is that
in QFTs which are weakly coupled
we can calculate and tell that there is not a large extra dimension
sticking out.
The fact that certain strongly coupled field theories
can be described by classical gravity on some auxiliary
spacetime is extremely powerful, once we believe it.

\subsection{Confidence-building measures}

Before proceeding to apply this machinery to
non-Fermi liquid metals,
we pause here to explain the reasons that give us
enough confidence in these \DV{rather odd} statements
to try to use them to do physics.
The reasons fall roughly into three categories

\begin{enumerate}
\item {\bf Many} detailed checks have been performed in special examples.
These checks have been done mainly in relativistic gauge theories
(where the fields are $N\times N$ matrices)
with extra symmetries (conformal invariance and supersymmetry).
The checks involve so-called `BPS quantities'
(which are the same at weak coupling and strong coupling or which
can be computed as a function of the coupling),
integrable techniques, and more recently some numerics.
We will not discuss any of these kinds of checks here, because they
involve calculations of quantities which only exist in specific models,
which neither the quantities nor the models are of interest here.

\item The holographic correspondence unfailingly gives
sensible answers for physics questions.
This includes rediscoveries of many known physical phenomena,
some of which are quite hard to describe otherwise:
\eg\ color confinement, chiral symmetry breaking, thermo, hydro, thermal screening, entanglement entropy, chiral anomalies, superconductivity, ...
%Nernst effect, fluctuation-dissipation theorem...}

The gravity limit, when valid, says who are the correct variables,
and gives immediate answers to questions about thermodynamics, transport, RG flow, ...
in terms of geometric objects.

\item If we are bold, the third class of reasons can be called ``experimental checks".
These have arisen from applications to the quark-gluon plasma (QGP)
produced at RHIC.
Holographic calculations have provided a benchmark value
for the viscosity of such a strongly-interacting plasma,
and have provided insight into the behavior of
hard probes of the medium, and of the approach to equilibrium.
%\HL{removed a couple of sociological sentences here.}

\end{enumerate}

As an illustration of the manner in which the correspondence
solves hard problems by simple pictures
we offer the following.
%Simple pictures for hard problems, an example}
The bulk geometry is a spectrograph separating the theory by energy scales.
The geometry dual to a general  Poincar\'e-invariant state of a QFT takes the form
\be
ds^2 = w(r)^2 \( \textcolor{darkgreen}{- dt^2 + d\vec x^2} \)  + R^2 {dr^2 \over r^2} .\ee
For the gravity dual of a CFT, the bulk geometry goes on forever, and the `warp factor' $w(r) = {r \over R} \to 0$.
%\begin{figure}
%\hskip0.5in
%\vskip-.2in
%\includegraphics[height=40pt]{pictures/blocking2.eps} % requires the graphicx package
%%$\longrightarrow$
%\hskip0.3in
%\includegraphics[height=60pt]{pictures/lance_figure1.eps} % requires the graphicx package
%\end{figure}
On the other hand, in the gravity dual of a model with a gap, the geometry ends smoothly, warp factor $w(r)$ has a nonzero minimum value.
If the IR region of the geometry is missing,
there are no low-energy excitations and hence an energy gap in the dual field theory.
\begin{figure}
\begin{center}
%\vskip-.1in
\includegraphics[height=130pt]{lance_figure_d.eps} % requires the graphicx package
\hskip.5in
\includegraphics[height=130pt]{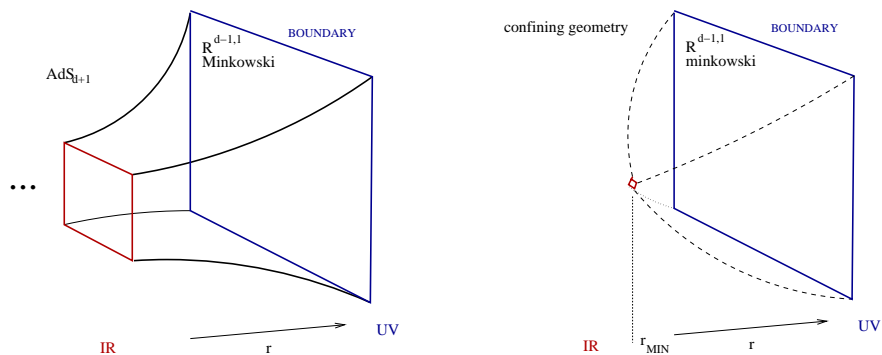} % requires the graphicx package
\caption{A comparison of the geometries associated with a CFT (left), and
with a system with a mass gap (right).}
\end{center}
\end{figure}

\subsection{Finite density}

A basic question for the holographic descripion is how to describe a finite density.
To approach this question, we introduce a minimal set of
necessary ingredients in the bulk model.
The fact that any local QFT has a stress tensor $T^{\mu\nu}$
means that we should have a dynamical metric $g_{\mu\nu}$ in bulk.
As a proxy for the conserved fermion number,
we assume that our QFT has a global abelian symmetry associated with a current $j^\mu$;
this implies the presence of a massless gauge field $A_\mu$ in the bulk.
As a proxy for bare electrons, we will also assume that our QFT
includes a fermionic operator $\Psi$;
this leads us to introduce to a bulk spinor field $\psi$.

Consider any relativistic CFT with a gravity dual and a conserved $U(1)$ symmetry.  %\redcom{$\exists$ many examples.}
The discussion goes through for any $d > 1+1$, but we focus on $d=2+1$.
The gravity dual is described by:
\be
\label{bulkaction}
  S = {1\over 2 \kappa^2} \int d^{4}x \sqrt g \( {{\cal R}} +
{ 6 \over R^2}  -
{ 2 \kappa^2 \over g_F^2} F_{\mu\nu}F^{\mu\nu} + ...\) \ee
The ellipsis indicates fields which vanish in groundstate, and more irrelevant couplings.
This is the action we would guess based on Wilsonian naturalness
and it's what comes from string theory when we can compute it.

As a warmup, let's discuss the holographic description of finite temperature.
The canonical ensemble of a QFT at temperature $T$ is described by
putting the QFT
on a space with periodic Euclidean time of radius $T^{-1}$.  The spacetime
on which the dual QFT lives is the boundary of the bulk geometry at $r \to\infty$.
So, the bulk description
of finite temperature is
the extremum of the bulk action whose Euclidean time
has a period which approaches $T^{-1}$ at the boundary.
%the saddle point
%of \eqref{GKPW}
For many bulk actions, including \eqref{bulkactionwithoutgaugefield},
this saddle point is a black hole in $AdS$.
This is a beautiful application of Hawking's observation that
black holes have a temperature.
%\begin{flushright}\vskip-.2in%\hspace{.75 in}
\begin{figure}[h!]
\begin{center}
\includegraphics[height=130pt]{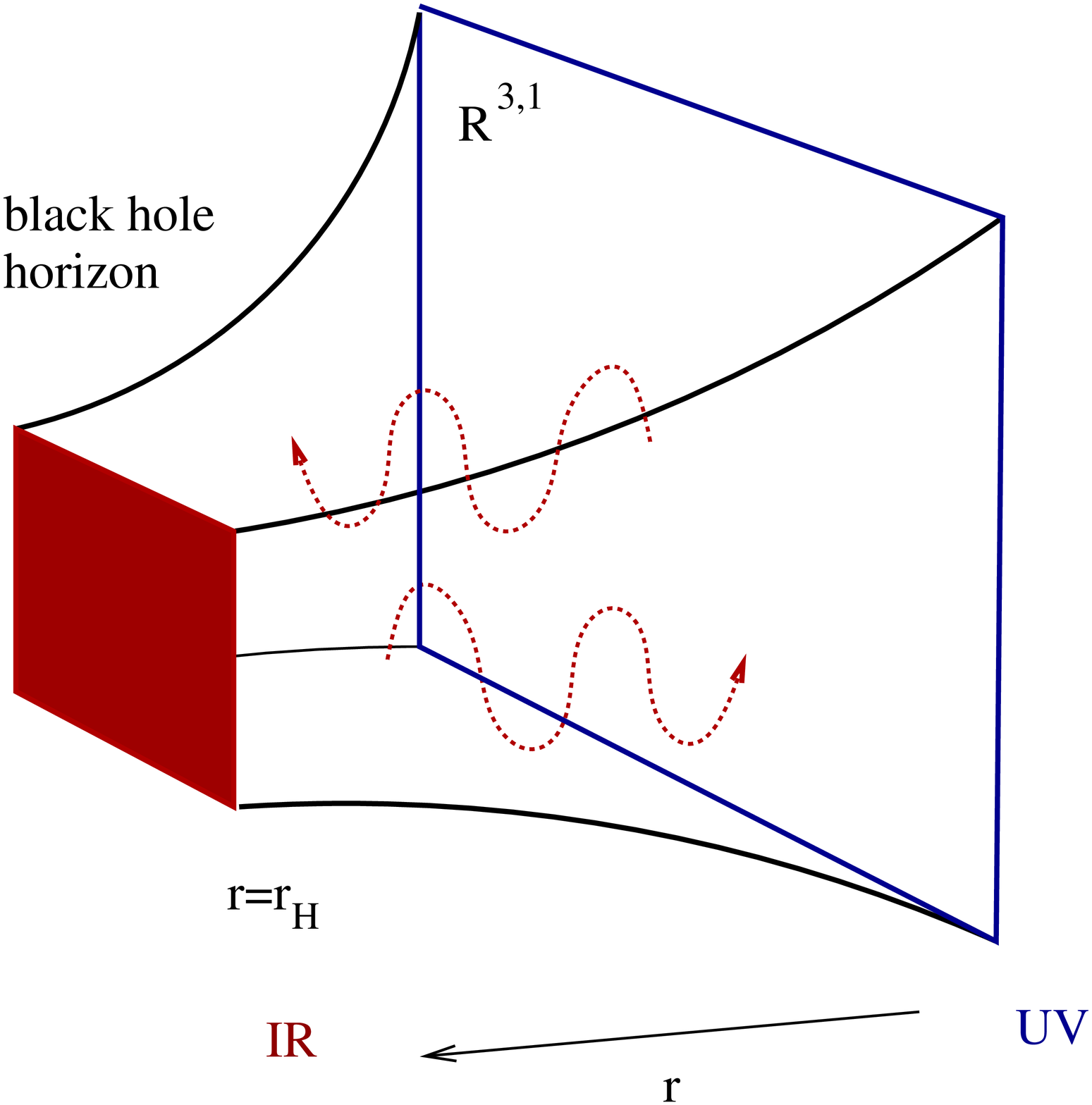} % requires the graphicx package
\end{center}
\end{figure}
%\end{flushright}

Similarly, the grand canonical ensemble is described by
periodic Euclidean time with a Wilson line
$e^{ i q \oint_\CC A^{(0)}} = e^{ - q \beta \mu }$
where $\CC$ is the thermal circle, and $A^{(0)}$ is the background
gauge field coupling to the current in question.
The source $A^{(0)}$ is the boundary value of the bulk gauge field.
For the minimal bulk field content introduced above,
the bulk solution with these boundary conditions
is the Reissner-N\"ordstrom (RN) black hole in $AdS$.
We will comment below on how this conclusion is modified
if we include other bulk fields, such as scalars, in \eqref{bulkaction}.

For zero temperature, the solution
of \eqref{bulkaction}
describing a finite density of $U(1)$ charge is
%$$ {ds^2\over L^2}  = {\alpha^2 \over z^2} \( - f^2(z) dt^2 + d\vec x^2 \)  + {dz^2\over z^2 h(z) } ,~~~~  A = q ( z-1)\alpha dt $$
%$$ f(z) = \sqrt{ 1 + q^2 z^4 - (1+q^2) z^3} $$
\be
\label{eq:RNBH}
 {ds^2}  = {r^2 \over R^2} \( - f dt^2 + d\vec x^2 \)  + R^2 {dr^2\over r^2 f } ,~~~~
%A = \mu \( 1 - \({r_0\over r} \)^{d-2} \) dt $$
A = \mu \( 1 - {1\over r}  \) dt \ee
where $ f(r) = 1 + {3 \over r^{4}} - {4\over r^3} $ is the zero-temperature limit.
This geometry has a horizon at $r=1$.
$\mu$ is the chemical potential for the $U(1)$ symmetry.

\subsection{Strategy to find a Fermi surface}

%\small
To look for a Fermi surface, we look for sharp features  in
fermionic Green's functions
at finite momentum and small frequency, following \cite{Lee:2008xf}.
%probe this system by measuring the response to bosonic and fermionic
%gauge-invariant operators.\\
%\ie\ we'll compute retarded two-point functions of the operators coupling to
%charged scalars and charged spinors.
%\greencom{This business was initiated by \ttref{Sung-Sik Lee, 0809.....}.}
Assume that amongst the $\dots$ in the bulk action \eqref{bulkaction} is\footnote
{
The derivative $\Dslash = \Gamma^MD_M$ contains the coupling to both
the spin connection and the gauge field
$ D_M \equiv \partial_M + {1\over 4} \omega_{MAB}\Gamma^{AB} - i q_\spinor A_M$.
}
%$$ S_{\rm probe}[\Phi] = \int d^{D+1}x \sqrt g \( | D\Phi |^2 + m^2 |\Phi|^2  + {\rm interactions} \) $$
\be S_{\rm probe}[\psi] = \int d^{D+1}x \sqrt g \( i \bar \psi \(\Dslash -
\textcolor{darkblue}{m} \) \psi    + {\rm interactions} \) \qquad 
\ee
The dimension and charge of the boundary fermion operator are determined by
\be \De = {d \ov 2} + \textcolor{darkblue}{m} R  , ~~~ \textcolor{darkblue}{q} = q.\ee
%the $\pm$ indicates an interesting ambiguity \cite{Klebanov:1999tb}
%which allows us to use the same bulk theory to describe two different QFTs
%by choosing different boundary conditions.
Here we can see a kind of `bulk universality': for two-point functions, the interaction terms don't matter.
We can describe many CFTs (many universality classes!) by a single bulk theory.
The results only depend on $q, \Delta$.
Some comments on the strategy:

\begin{itemize}

\item There are many string theory vacua with these ingredients \cite{modulistabilization1,modulistabilization2,modulistabilization3,modulistabilization4,modulistabilization5}.
In specific examples of dual pairs
(\eg\ M2-branes $\leftrightsquigarrow$ M-theory on $AdS_4 \times S^7$),
the interactions and the parameters
$ q, m  $ are specified.
Which sets $ \{ q, m \} $ are possible
and what correlations there are is not clear,
and, as an expedient, we treat them as parameters.

\item This is a large complicated system (with a density $\rho \sim N^2$),
of which we are probing a tiny part (the fermion density can be seen to scale
like $\rho_\Psi \sim N^0$).

%\item
%To the extent that a weakly-curved gravity theory requires a strongly coupled
%dual field theory,
%it would be surprising if we could describe a Fermi liquid (a weakly coupled QFT),
%using these techniques.

%

\item In general,
both bosons and fermions of the dual field theory
are charged under the $U(1)$ current: this is a Bose-Fermi mixture.
The relative density of bosons and fermions, and whether
the bosons will condense, is a complicated dynamical
question.  Fortunately, the gravity theory solves this
problem for us; let's see what happens.

\end{itemize}

%[PROCEED IN A SPIRIT OF EMPIRICISM.]

\subsection{AdS/CFT prescription for spinors}

To compute the retarded single-fermion Green's function $G_R$, we must solve the
Dirac equation $(\Dslash + m)\psi=0$ in the black hole geometry, and
impose infalling boundary conditions at the horizon
\cite{Son:2002sd, Iqbal:2009fd}.  Like retarded response, falling into the black hole is something that {\it happens}, rather than unhappens.
%\DV{(unhappens?)}.
Translation invariance in $\vec x, t$ turns the Dirac equation into an ordinary differential equation in $r$.
Rotation invariance allows us to set $k_i = \delta_i^1 k$;
we can then choose a basis of gamma matrices in which the Dirac equation
is block diagonal and real\footnote{It is also convenient to redefine the independent variable
by $\psi = \( - \det g g^{rr}\)^{-1/4}\Phi$.}.
Near the boundary, solutions behave in this basis as
\be
\label{eqn:bc}
\Phi {\buildrel{ r \to \infty}\over{\approx}}  a_\alpha r^{m}
\left( \begin{matrix} 0 \cr 1 \end{matrix} \right)
+  b_\alpha r^{-m}
\left( \begin{matrix}1 \cr 0 \end{matrix} \right)~~~.
\ee
From this data, we can extract a matrix of Green's functions, which has two independent eigenvalues:
\be G_\alpha(\omega, \vec k) = { b_\alpha \over a_\alpha}, ~~ \alpha = 1,2\ee
The label $\alpha=1,2$ indexes
a multiplicity which arises in the boundary theory
as a consequence of the short-distance Lorentz invariance and
will not be important for our purposes.
The equation depends on $q$ and $\mu$ only through %the effective chemical potential
$
 \mu_q \equiv \mu q.
$
We emphasize that all frequencies
%In our gauge $A_t(r_0) = 0 , A_t(r\to\infty) = \mu $:
%$ u \buildrel{r\to\infty}\over{\approx} \omega + \mu_q $.\\
$\omega$
appearing below are measured from the effective chemical potential, $\mu_q$.
All dimensionful quantities below are quoted in units of the chemical potential.

\section{A Fermi surface}

The system is rotation invariant, $G$ depends on $k = |\vec k|$ and $\omega$. \DV{The Green's function satisfies
the following:}

%We find a pole in $G_{2}(\omega)$ \greencom{($m=0, q=1\; (\mu_q = \sqrt{3})$)}\\
%for $k<k_F=0.9185 \pm 0.0001$, which sharpens
%and passes through zero frequency from below at $ k=k_F$.

\begin{figure}[h!]
\begin{center}
%\vskip-0.15in
%\hfill
%\hskip-1in
%\includegraphics[totalheight=3.4cm,origin=c,angle=0]{envelope_finite_both_q1.eps}
\includegraphics[totalheight=4.6cm,origin=c,angle=0]{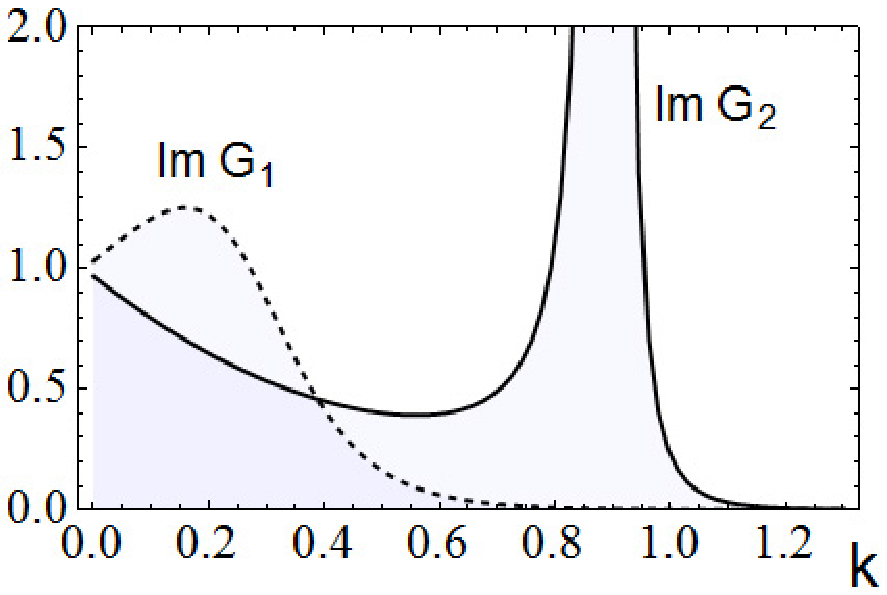}
~~~~
\includegraphics[scale=0.5,origin=c]{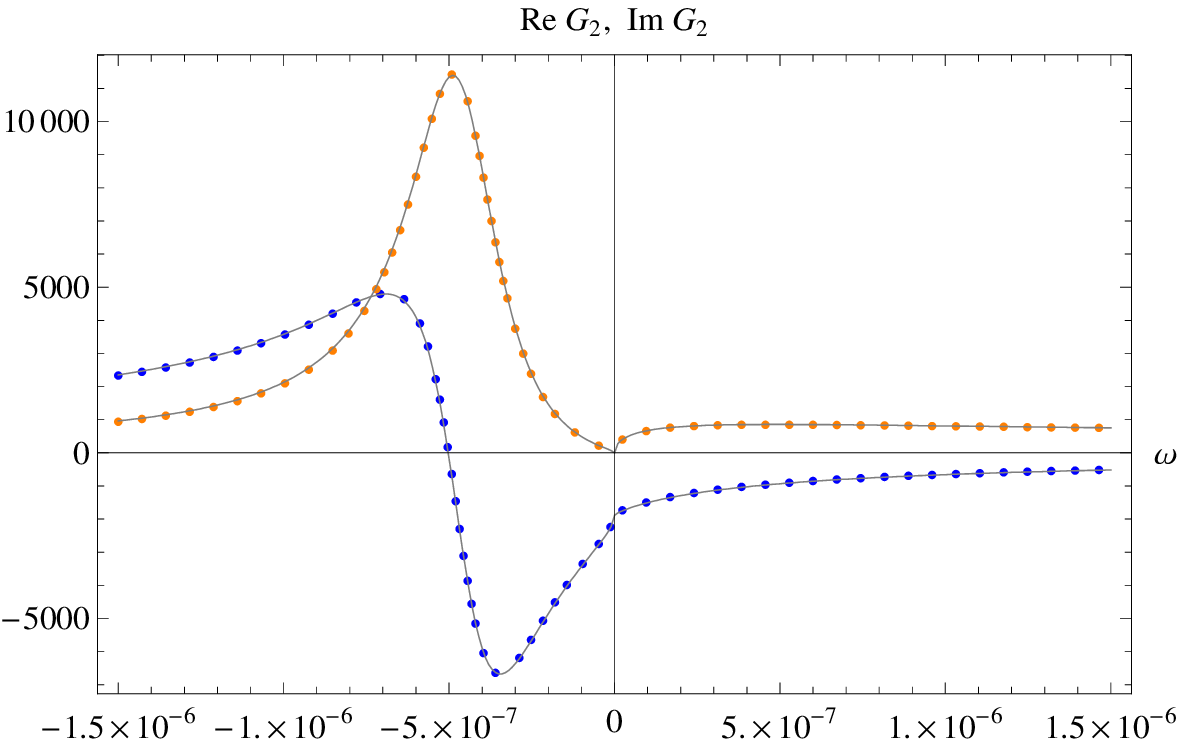}
%\includegraphics[scale=0.39]{pictures/green_function_12.eps}
%~~~~
%\includegraphics[scale=0.4]{pictures/green_function_3.eps}
%\hskip-0.1in\hfill
%\caption{\label{fig:Sp} Spectral function $\Im G_{22}(\omega)$ at $k=1.2 < \mu_q$ (left plot) and $k=3.0 > \mu_q$ (right plot) for $m=0$ and $q=1\; (\mu_q = \sqrt{3})$. The function asymptotes to $1$ as $|\om| \to \infty$ as in the vacuum~\eqref{vac}. Right plot: The onset of the finite peak  at $\omega \approx 1.2 \approx k - \mu_q$ roughly corresponds to the location of divergence at $\omega = k$ in the vacuum~\eqref{vac}. The function is roughly zero between $\omega \in (-k - \mu_q, k-\mu_q)$ again recovering the behavior in the vacuum.
%Left plot: The deviation from the vacuum behavior becomes significant.
%}
\\
MDC: \ $\Im G_{1,2}(\omega = -10^{-3}, k)$\hskip0.6in EDC: \ $G_2(\omega, k = 0.9)$
\vskip0.1in
$\greencom{For~q=1, m=0:}~~~ k_F \approx 0.918528499 $
\end{center}
\end{figure}
%\vskip-0.2in
%$$k_F \approx 0.918528499 $$
%Left: $k=1.2 < \mu_q$, ~ Right: $k=3.0 > \mu_q$.

%Here are some basic checks on this result:
\begin{itemize}
\item{} The spectral density is positive for all $\omega$ as required by unitarity.

\item{} The only non-analyticity in $G_R$ (for real $\omega$) occurs at $\omega=0$.
This could be anticipated on physical grounds, and is a consequence of the following argument.
Consider a mode that is normalizable at the boundary, \ie\ $a = 0$ in \eqref{eqn:bc}.
%\be
%\phi(r) \sim 0 \times r^{d-\Delta} + ...
%\ee
This is a real boundary condition. Thus if the equations of motion are real (which they are for real $\om$), the wave must be real at the other end, {\it i.e.} at the horizon. Thus it must be a sum of both infalling and outgoing waves at the horizon.
Thus it can't be an infalling solution.
The exception is at $\om = 0$ where the infalling and outgoing conditions are each real.

\item The result approaches the CFT behavior at higher energies ($|\omega| \gg \mu$).
\end{itemize}

At $T=0$, we find numerically \DV{a quasiparticle peak near $k=k_F \approx 0.9185$. }
The peak moves with dispersion relation $\omega \sim k_\perp^z$
with
$ z = 2.09$ for $q=1, \Delta = 3/2$
and
$ z = 5.32$ for $q=0.6, \Delta = 3/2$.
There is scaling behavior near
the Fermi surface
\be G_R(\lambda k_\perp, \lambda^z \omega)
= \lambda^{-\alpha} G_R(k_\perp, \omega) , ~~~ \alpha = 1 .\ee
It's not a Fermi liquid. \DV{The above} scaling should be contrasted with the scaling in a Landau Fermi liquid, $z = \alpha = 1$. Finally, in the cases shown here, the residue vanishes at the Fermi surface.

\begin{figure}[h] \begin{center}
\includegraphics[scale=0.7,angle=0]{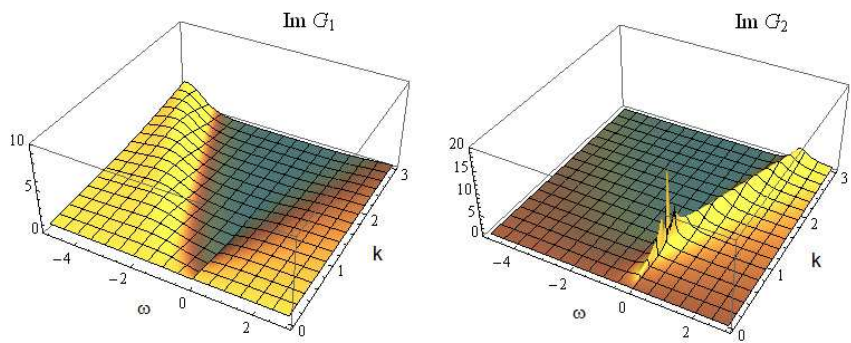}
\caption{\label{fig:three_d}
3d plots of $\Im G_{1}(\omega, k)$ and $\Im G_{2}(\omega, k)$ for $m=0$ and $q=1 \, (\mu_q = \sqrt{3})$. In the right plot the ridge at $k \gg \mu_q$ corresponds to the smoothed-out peaks at finite density of the divergence at $\om = k$ in the vacuum. As one decreases $k$ to a value $k_F \approx 0.92 < \mu_q$, the ridge develops into an (infinitely) sharp peak indicative of a Fermi surface.
}
%\caption{\label{fig:three_d}
%3d plots of $\Im G_{11}(\omega, k)$ and $\Im G_{22}(\omega, k)$ for $m=0$ and $q=1 \, (\mu_q = \sqrt{3})$. In the right plot the ridge at $k \gg \mu_q$ corresponds to the smoothed-out peaks at finite density of the divergence at $\om = k$ in the vacuum. As one decreases $k$ to a value $k_F \approx 0.92 < \mu_q$, the ridge develops into an (infinitely) sharp peak indicative of a Fermi surface.
%}
\end{center}
\end{figure}

\section{Low-frequency behavior}

So far, AdS/CFT is a black box which produces
consistent spectral functions.
To understand better where these numbers
come from, we need to
take apart the black box a bit.

\subsection{Emergent quantum criticality from geometry}

%The
% \be {ds^2}  = {r^2 \over R^2} \( - f dt^2 + d\vec x^2 \)  + R^2 {dr^2\over r^2 f }  \ee
At $T=0$,
the `emblackening factor'
in \eqref{eq:RNBH} behaves as $f \approx 6 (r-1)^2 $
and this means that the
near-horizon geometry is $AdS_2 \times \RR^{d-1}$.
Recall that $AdS$ means conformal symmetry.
The conformal invariance of this metric is {\it emergent}.
We broke the microscopic conformal invariance
when we put in the finite density.
%\begin{figure}[h] \begin{center}
%\includegraphics[scale=0.35]{pictures/extremalBH1.eps}
%\end{center}
%\end{figure}
%\vskip-.2in
%\hskip1.1in$\omega \ll \mu$ \hskip 1.25in $\omega \gg \mu$\\
AdS/CFT suggests that the low-energy physics is governed by
the dual {\it IR CFT}.
More precisely, there is such a ${\rm CFT}_0$
%\redcom{(or chiral ${\rm CFT}_1$)}
for each $\vec k$.
At small temperature $T \ll \mu$,
the geometry is a black hole in $AdS_2$ times the space directions.
The bulk geometry is a picture of the RG flow from the CFT${}_d$ to this non-relativistic CFT.
\begin{figure}[h]
\begin{center}
\includegraphics[scale=0.35]{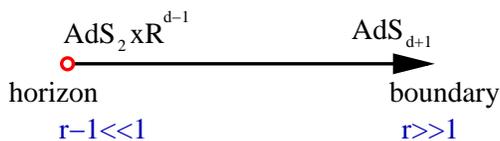}
\caption{\label{extremalBH}
%\hskip1.3in$\omega \ll \mu$ \hskip 1.25in $\omega \gg \mu$
The geometry of the extremal $AdS_{d+1}$ charged black hole.
}
\end{center}
\end{figure}
%\vskip-.2in

\subsection{Matching calculation}

At high temperatures
({\it i.~e.} not small compared to the chemical potential)
the retarded Green's function
 $G_R(\omega)$ is analytic near $\omega=0$, and
 can therefore be conveniently computed in series expansion \cite{Son:2002sd}.
In the limit of interest to us, $T \ll \mu$,
expanding the wave equation in $\omega$ is delicate.
This is because the $\omega$-term dominates near the horizon.
We proceed using the method of matched asymptotic expansions:
we find the solution (in an $\omega$-expansion) in two regions of BH geometry (IR and UV),
and fix integration constants by matching their behavior in
the region of overlap.
The region of overlap is large when $\omega, T \ll \mu$.
This technique
is familiar to string theorists
from the brane absorption calculations \cite{magoo} which
led to the discovery of holographic duality.
Here,
this `matching' can be interpreted in the QFT as RG matching
between UV and IR CFTs.
Note that we need only assume the existence of the IR CFT; the gravity dual lets us compute.

First we discuss the IR boundary condition;
this is associated with the near-horizon region of the bulk geometry,
which is $AdS_2 \times \RR^2$.
Wave equations for charged fields in $AdS_2$ turn out to be solvable.
Near the boundary of $AdS_2$, the solutions for the mode with momentum $k$
are power laws with exponents $\pm \nu_k$ where\footnote{
In the case of $d-1$ space dimensions, the formula becomes
$ \nu = R_2 \sqrt{ m^2 + k^2 - q^2e_d^2 } $
where we define $ e_d \equiv {g_F\over \sqrt{2d(d-1)} } $.
}

%$$ \psi \buildrel{\sigma \to 0}\over {\approx} \sG \sigma^\nu v_+ + \sigma^{-\nu} v_-$$
%$$
%\yandz_{\alpha}^{(0) }(\sigma)
%\buildrel{\sigma \to 0}\over {\approx}
% U_\alpha
%\left( \begin{matrix} \sG \sigma^{\nu} \cr \sigma^{-\nu}  \end{matrix}\right)
%, ~~~
%\small{U_\alpha \equiv \left( \begin{matrix}  -m- {\nu \over R_2} & - m + {\nu \over R_2} \cr
%(-1)^\alpha k - \hat \mu_q & (-1)^\alpha k - \hat \mu_q  \end{matrix}\right)}~;
%$$
\be \nu_k \equiv R_2 \sqrt{ m^2 + k^2 - q^2/2 } ~~~.
\ee
This exponent determines the scaling dimension of the IR CFT operator
$\CO_k$
to which the spinor operator with momentum $k$ flows:
$ \delta_k = \half + \nu_k .$
%For a spinor in $AdS_2$, $k$ is a parity-violating mass term
%$\tilde m \bar \psi \Gamma \psi $:  $\tildem \equiv k {R_2 \over r_0}$ }
%
%$ \Psi(\omega, k) $ matches onto some IR CFT operator $\CO_k$ of dimension
%$\delta_k = \half + \nu_k$, \greencom{whose two-point function is the}
The retarded two-point function of the operator $\CO_k$
will play an important role, and
is
\be
\label{eq:sGdef}
\sG_k(\omega)
=  c(k) \omega^{2\nu_k} ~.
\ee
$c(k)$ is a complex function, whose calculation and explicit form is
described in next section.
%\be
%%\sG_R (\om) =
%c(k) = 2^{2\nu}
%e^{- i \pi \nu} \frac{\Gamma (-2 \nu ) \, \Gamma \left(1+\nu -i q e_d \right)}{\Gamma (2 \nu )\,   \Gamma \left(1-\nu -i q e_d\right)}
%% \frac{m R_2 - i q e_d - \nu}{ m R_2 - i q e_d +  \nu} \( 2\om\)^{2 \nu}.
%\cdot \frac{ \le(m + i \tildem \ri) R_2- i q e_d - \nu}
%{  \le(m + i \tildem \ri) R_2 - i q e_d +  \nu}
%%\( 2\om\)^{2 \nu}
%\ee

Next we consider the low-frequency expansion in the
near-boundary region of the geometry, which is associated with the UV of the field theory.
The outer region solutions can be expanded in powers of $\omega$,
and we can work in a basis where the Dirac equation is completely real.
The UV data is therefore real and analytic in $\omega$.
A basis of solutions at $\omega=0$ is
\be\label{eqn:nearUV}
\yandz_{\alpha}^{(0)\pm }
\buildrel{r \to 1} \over {\approx}
v_\pm (r-1)^{\mp \nu}
% U_\alpha \left( { 1 \mp \sigma^3 \over 2} \right)
%\left( \begin{matrix}  (r-1)^{-\nu} \cr (r-r_\star)^{\nu}  \end{matrix}\right)
\ee
where $v_\pm$ are certain constant spinors which will be specified in
Section \ref{sec:cor} below.
Two solutions can then be constructed perturbatively in $\omega$:
\be\yandz_{\alpha}^{\pm }=
\yandz_{\alpha}^{(0)\pm }
+ \omega \yandz_{\alpha}^{(1)\pm }
+ \omega^2 \yandz_{\alpha}^{(2)\pm } + \dots ,
~~~
\yandz_\alpha^{(n)\pm} \buildrel{r \to \infty}\over {\approx}
\left( \begin{matrix}  \textcolor{darkblue}{b_\alpha^{(n)\pm}} r^{-m} \cr
\textcolor{darkblue}{a_\alpha^{(n)\pm}} r^{m} \end{matrix}\right).
\ee
Matching these solutions to the leading and subleading solutions in the near-horizon region gives
\be
\psi_\alpha = \psi_{\alpha}^{+ } + \textcolor{darkred}{\sG (\omega)}
\psi_{\alpha}^{-}
\ee
where $\sG$ is the IR CFT Green's function defined above in \eqref{eq:sGdef}.
%$$
%\boxed{ G_{R} (\om, k) = K {{\bl{b_+^{(0)}} + \om \bl{b_+^{(1)}} + O(\om^2) +
%\textcolor{darkred}{\sG_k (\om)} \le( \bl{b_-^{(0)}} + \om \bl{b_-^{(1)}} + O(\om^2)\right)
%\ov \bl{a_+^{(0)}} + \om \bl{a_+^{(1)}} + O(\om^2) + \textcolor{darkred}{\sG_k (\om)}
% \le(\bl{a_-^{(0)}} + \om \bl{a_-^{(1)}} + O(\om^2)\right)
% }}}
%% \ov {
%% a_+^{(0)} + \om a_+^{(1)} + O(\om^2) + \textcolor{darkred}{\sG_k (\om)}
%% \le(a_-^{(0)} + \om a_-^{(1)} + O(\om^2)\right) } }}
%$$

For any $k$, this produces a Green's function of the form:
\be
\label{eq:mainresult}
\boxed{ G_{R} (\om, k) =
K {{\bl{b_+^{(0)}} + \om \bl{b_+^{(1)}} + O(\om^2) +
\textcolor{darkred}{\sG_k (\om)} \le( \bl{b_-^{(0)}} + \om \bl{b_-^{(1)}} + O(\om^2)\right)
\ov \bl{a_+^{(0)}} + \om \bl{a_+^{(1)}} + O(\om^2) + \textcolor{darkred}{\sG_k (\om)}
 \le(\bl{a_-^{(0)}} + \om \bl{a_-^{(1)}} + O(\om^2)\right)
 }}}
\ee
\DV{The overall factor $K$ will not be important in the following and we set it to unity.}
For generic $k$, the UV coefficient $\bl{a_+^{(0)}(k)}$ is nonzero.  The low-frequency expansion gives
\be G_{R} (\om, k) = \bl{{b_+^{(0)} \over a_+^{(0)} }} + \bl{r_1} \omega +
\bl{r_2} \textcolor{darkred}{\sG_k(\omega)} + \dots\ee
%$$ \redcom{\sG_k(\omega) = c(k) \omega^{2\nu_k} }$$
Since $a_\pm, b_\pm$ are all real,
we see that the nonanalytic behavior and dissipation are controlled by IR CFT.

\subsection{Consequences for Fermi surface}
%$$
%\boxed{
%G_{R} (\om, k) =
%K {{\bl{b_+^{(0)}} + \om \bl{b_+^{(1)}} + O(\om^2) +
%\textcolor{darkred}{\sG_k (\om)} \le( \bl{b_-^{(0)}} + \om \bl{b_-^{(1)}} + O(\om^2)\right)
%\ov \bl{a_+^{(0)}} + \om \bl{a_+^{(1)}} + O(\om^2) + \textcolor{darkred}{\sG_k (\om)}
% \le(\bl{a_-^{(0)}} + \om \bl{a_-^{(1)}} + O(\om^2)\right)
% }}}
%$$
Suppose there is some $k = k_F$ such that $\bl{a_+^{(0)}(k_F)} = 0$
in \eqref{eq:mainresult}.
This happens if there exists a zero-energy boundstate of outer-region Dirac equation;
it could be called `inhomogeneous fermionic hair' on the black hole.
Near such a value of $k$,
%
%\end{frame}
%%\subsection{Implications}
%\begin{frame}{Consequences for Fermi surface}
\be
\label{eq:GR}
 G_R(\omega, k ) = { \bl{h_1} \over k_\perp - {1\over \bl{ v_F}}  \omega - \bl{h_2 }\textcolor{darkred}{c(k)} \omega^{2 \textcolor{darkred}{\nu}_{k_F}} } .
\ee
The coefficients $\bl{h_{1,2}, v_F}$ are real, UV data.
%\vskip-.52in
%\greencom{(if $2\nu_{k_F} \in \IZ$: $ \omega^{2\nu}  \to \omega^{2 \nu}\log \omega $)}
This form of the Green's function correctly fits numerics near the Fermi surface.
%\begin{figure}[h!]
% \begin{center}
% \hskip0.5in
%\includegraphics[scale=0.31]{pictures/fits1.eps}
%\caption{
%Numerical results for
%$ k = .9, m=0, q=1 \Longrightarrow \nu_{k_F} =
%{1\over \sqrt{6} } \sqrt{ m^2 + k_F^2 - q^2/2 }  = 1.045... > \half $.
%Also plotted is the function \eqref{eq:GR}.
%}
%\end{center}
%\end{figure}
The expression \eqref{eq:GR} contains a lot of information
about the behavior near the Fermi surface.
The result depends on the value of $\nu$ compared to $\half$.

%\begin{itemize}
%\item{$\nu<\half$: singular Fermi liquid}
%\nu_{k_F} \equiv \sqrt{ m^2 + k_F^2 - q^2/2 }/\sqrt{6}
First suppose \textcolor{darkred}{$\half > \nu_{k_F}  \equiv \sqrt{ m^2 + k_F^2 - q^2/2 }/\sqrt{6}$}.
This means that
the IR CFT operator $ \textcolor{darkred}\CO_{k_F}$ is relevant,
\ie\ $\delta_{k_F} = \half + \nu_{k_F} < 1$; we will explain the significance of this in the final section.
This means that the non-analytic term dominates near $\omega=0$:
\be G_R(\omega, k ) = { h_1 \over k_\perp + {1\over v_F}  \omega + c_k \omega^{2\nu_{k_F}} }
~~~
%
%$$ G_R(\omega, k ) = { h_1 \over k_\perp - {1\over v_F}  \omega - h_2 \omega^{2\nu_{k_F}} }
%$$
\omega_\star(k) \sim k_\perp^z, ~~~~~ z = { 1\over 2 \nu_{k_F}} > 1 \ee
The excitation at the Fermi surface is not a stable quasiparticle:
\be {\Gamma(k)\over \omega_\star(k) }
\buildrel{ k_\perp \to 0} \over {\to} {\rm const},
~~~~~~~ Z \propto k_\perp^{ { 1 - 2 \nu_{k_F}} \over 2 \nu_{k_F}}
\buildrel{ k_\perp \to 0} \over {\to}  0. \ee

%\item{$\nu>\half$: Fermi liquid}
Suppose $\textcolor{darkred}{\nu_{k_F} >  \half}$;
in this case
$\textcolor{darkred}\CO_{k_F}$ is irrelevant and the linear term dominates the dispersion:
%~~~~$\delta_k = \half + \nu_k > 1 $.
\be G_R(\omega, k ) = { h_1  \over k_\perp + {1\over v_F}  \omega + c_k \omega^{2\nu_{k_F}} }
~~~
 \omega_\star(k) \sim v_F k_\perp \ee
There is a stable quasiparticle, but
%\redcom{
the system is not Landau Fermi liquid in that the lifetime goes like a funny power of frequency:
%} (different thermo, transport.)
\be {\Gamma(k)\over \omega_\star(k) }  \propto k_\perp^{ 2 \nu_{k_F} - 1 } \buildrel{ k_\perp \to 0} \over {\to} 0
~~~~~~ Z
\buildrel{ k_\perp \to 0} \over {\to} h_1 v_F. \ee

%\begin{figure}[h!]
% \begin{center}
%\includegraphics[scale=0.60]{pictures/poles2-nonfl.eps}
%~~~
%\includegraphics[scale=0.40]{pictures/poles3-fl.eps}
%\end{center}
%\caption{The motion of the pole in the complex frequency plane as $|\vec k|$ is varied through $k_F$, for
%$\nu < \half$ (left) and $\nu > \half$ (right).}
%\end{figure}
%

%\item{$\nu=\half$: Marginal Fermi liquid}
%\end{frame}\begin{frame}{Marginal operator: ``Marginal Fermi liquid"}

Finally, suppose $\nu_{k_F} =  \half$: $\CO_{k_F}$ is {\it marginal.}
%~~~~$\delta_k = \half + \nu_k=1$.\\
%$$ G_R(\omega, k ) = { Z \over k_\perp + {1\over v_F}  \omega + c_k \omega^{2\nu_{k_F}} } $$
%
%$$ G_R(\omega, k ) = { h_1 \over k_\perp - {1\over v_F}  \omega - h_2 \omega^{2\nu_{k_F}} }
%$$
The two frequency-dependent terms compete. $v_F \propto \nu_{k_F} - \half \to 0$, while $c(k_F)$ has a pole;
they cancel and leave behind a logarithm:
\be G_R \approx { h_1 \over k_\perp + \tilde c_1 \omega \ln \omega + c_1 \omega },
~~~ \tilde c_1\in \RR, ~~ c_1 \in \IC
~~~~
Z \sim { 1\over | \ln \omega_\star| }
\buildrel{ k_\perp \to 0} \over {\to} 0    .
\ee
This is the form of the Green's function
proposed in a well-named phenomenological model of the electronic excitations of the cuprates near optimal doping
\cite{varma}.
%\ttref{[Varma et al, 1989]}.}
%\includegraphics[scale=0.30]{pictures/mfl.eps}

The case $\nu_{k_F} = 1$ appears similar to a Landau Fermi liquid \cite{Cubrovic:2009ye},
though there are also logarithms in this case, and the
physical origin of the quasiparticle decay is quite distinct from
electron-electron interactions.

%\end{itemize}
\subsection{UV data: where are the Fermi sufaces?}

\begin{figure}[h!]
\begin{center}
\includegraphics[scale=0.70]{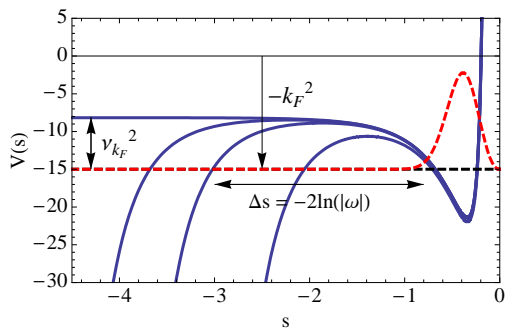}
\caption{\label{fig:pot}
Shown here is a sequence of Schr\"odinger potentials for the scalar wave equation
in the RN black hole.   The horizontal axis is a `tortoise' coordinate $s$ which
makes the wave equation into a one dimensional Schr\"odinger problem.
The role of the energy in the Schr\"odinger problem
is played by $-k^2$.
The red dotted curve is a cartoon of the boundstate wavefunction at $\omega=0$
with energy $-k_F^2$;
the blue curve which becomes horizontal at large negative $s$ (the IR region) is
the associated Schr\"odinger potential for $\omega=0$.
As $\omega$ increases from zero, the potential develops a well in
the IR region (the other blue curves), into which the boundstate can tunnel.
The width of the barrier is $\Delta s \sim -2\ln|\omega|$,
and the height is $\nu_{k_F}^2$; hence the tunneling
amplitude which determines the decay rate of the Fermi surface boundstate is
$ e^{ - \text{area}} \sim \omega^{2 \nu_{k_F} } $.
}
\end{center}
\end{figure}

Above we assumed \DV{ $a_+^{(0)}(k_F) = 0 $ }.  This happens at
$k=k_F$ such that there exists a normalizable, incoming solution at $\omega=0$.
From a relativist's viewpoint, the black hole acquires inhomogeneous fermionic hair.
By a change of variables, this problem can be translated into a boundstate problem in one-dimensional quantum mechanics.  The relevant Schr\"odinger potentials
are shown in Fig.~\ref{fig:pot}.
%\vskip-.2in
%\begin{figure}[h!]
%\begin{center}
%\includegraphics[scale=0.30]{potentials.eps}
%\caption{\label{fig:pot}
%Schr\"odinger potentials $V(\tau)/k^2$ at $\omega=0$ %or various values of $k$
%for $m<0, m=0, m>0$ respectively.
%%All pictures are of $V(\tau)/k^2$ with
%$\tau$ is the tortoise coordinate. %and $q/k$ varied. \\
%Right ($\tau=0$) is boundary; left is horizon.
%}
%\end{center}
%\end{figure}
A few cases are worth noting.
For $k > qe_d$, the potential is always positive and there can be no boundstate;
there is therefore no Fermi surface in this regime.
Below a certain momentum, $k < k_{osc} \equiv \sqrt{ (qe_d)^2 - m^2 }$,
the potential develops a singular well near the horizon
$V(x) \sim {\alpha\over \tau^2}$,
with $\alpha < - {1\over 4} $.
We refer to this as the ``oscillatory regime";
it is associated with particle production in the $AdS_2$ region of the geometry.
In this case, the exponent $\nu$ is imaginary, and hence $G_R$ periodic in log $\omega$.
We note that this behavior is quite independent of the
Fermi surface behavior.

\begin{figure}[h]
 \begin{center}
\includegraphics[scale=0.3]{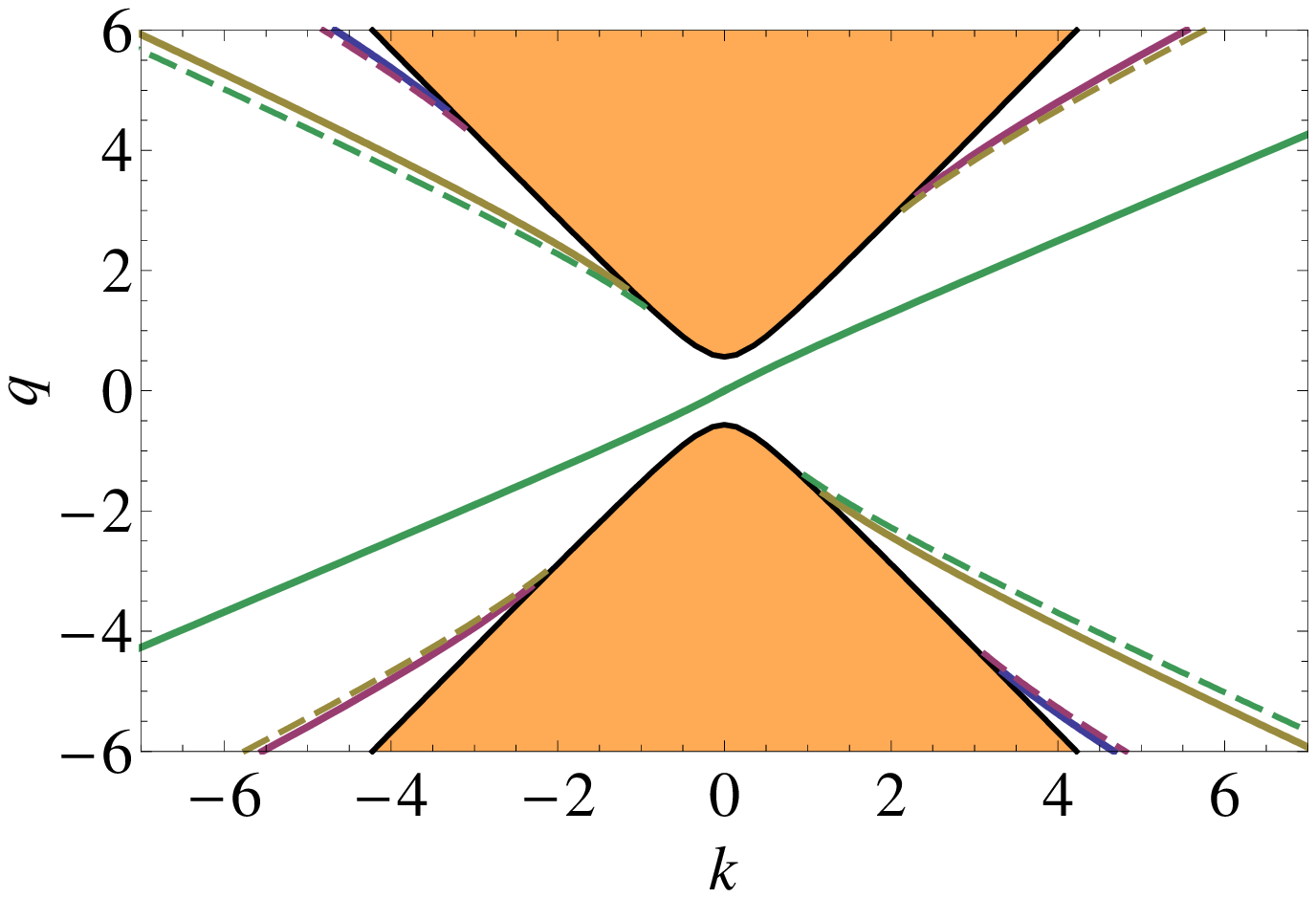}
\includegraphics[scale=0.3]{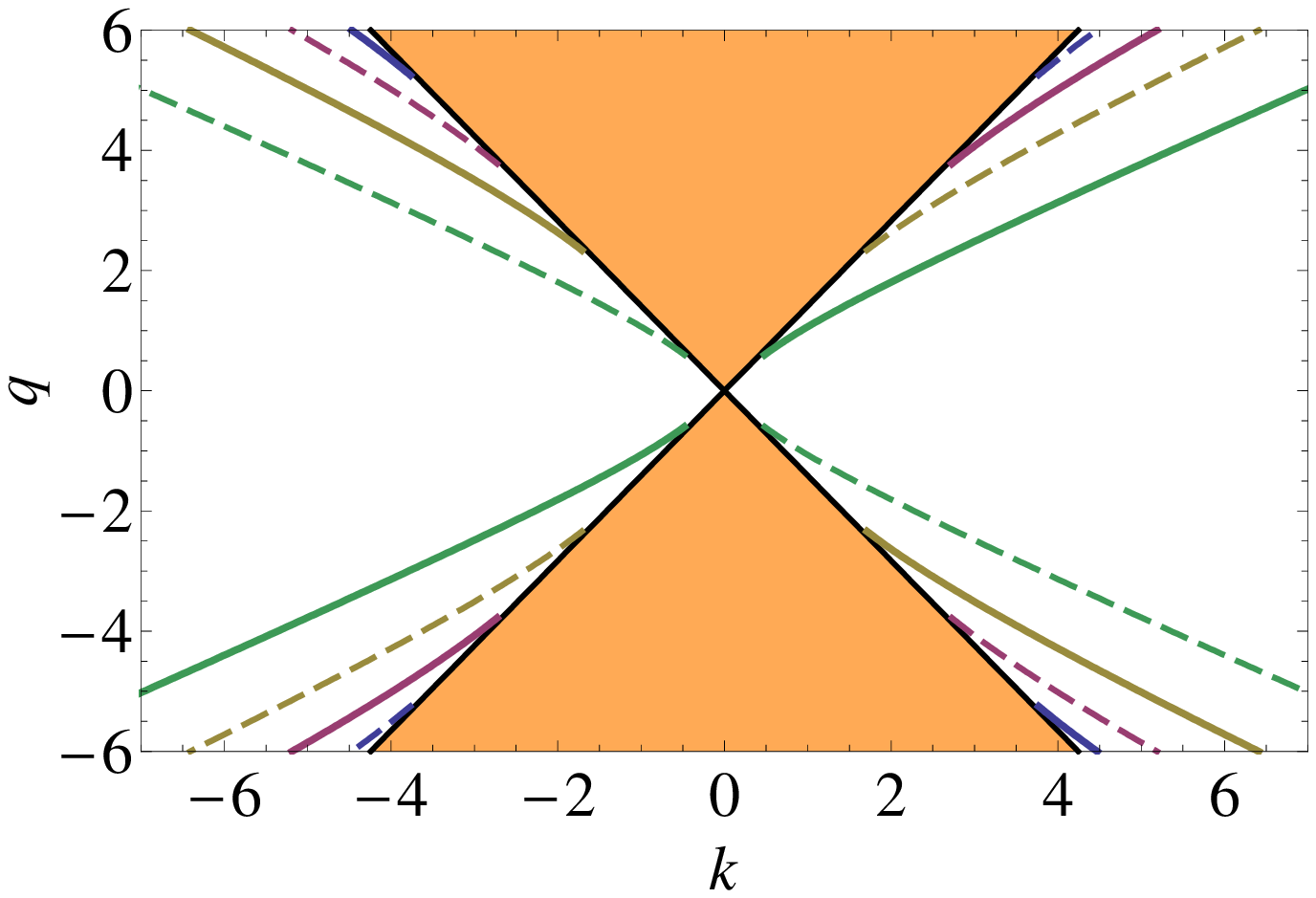}
\includegraphics[scale=0.3]{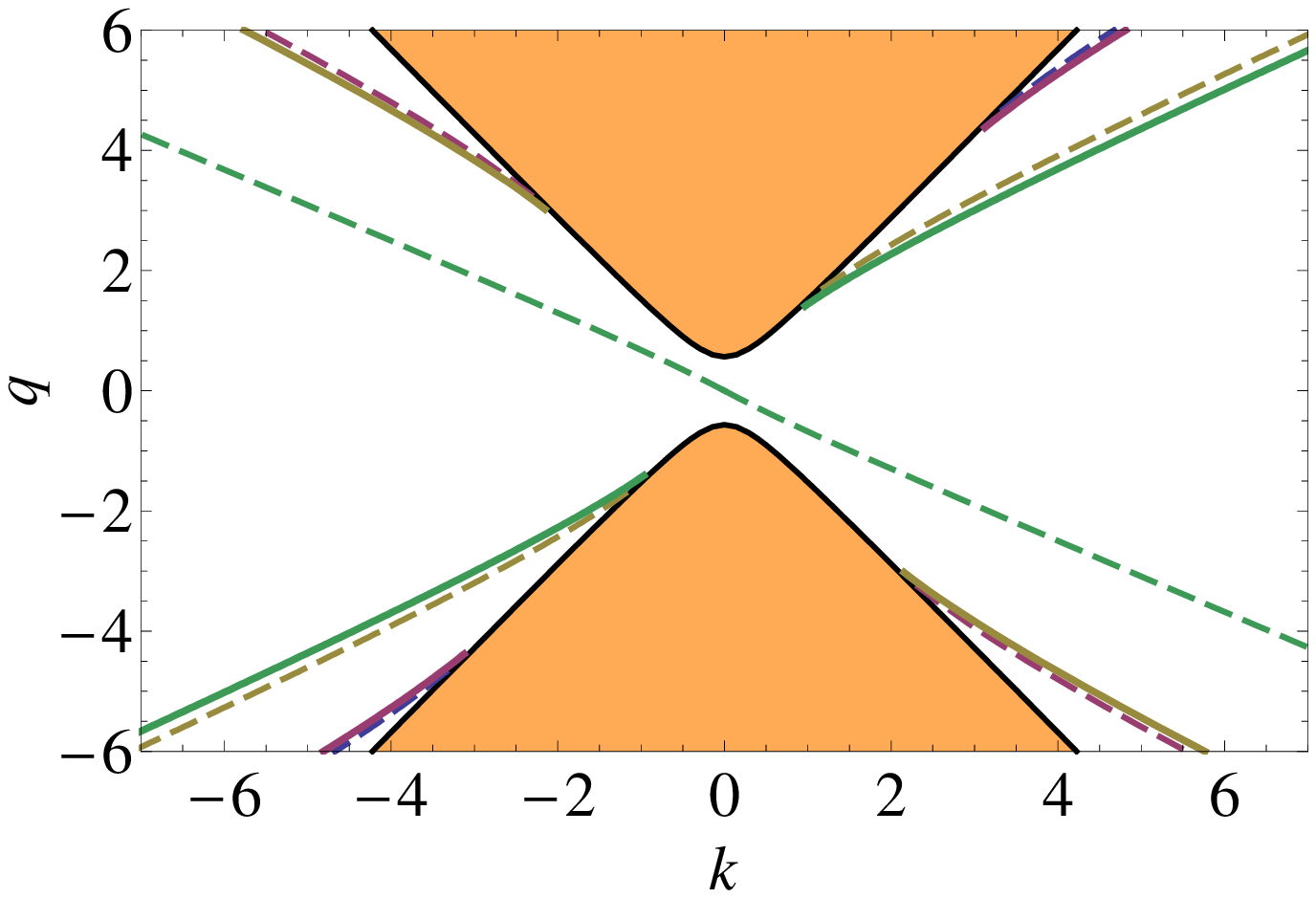}
\caption{\label{fig:qkplots}
The values of $q, k$ at which
poles of the Green's function $G_{2}$ occur
are shown by solid lines
for $m = -0.4, 0, .4$.
The dotted lines represent zeros of $G_{2}$,
and hence poles of the Green's function $\tilde G_{1}$
for the opposite sign of the mass.
The orange bits
indicate the oscillatory region, where $\nu \in i \IR$.
%for the alternative quantization.
%We observe that as $m \to - {1\over 2}$,
%a line of poles and a line of zeros of $G_{1}$ approach
%each other.
%At $m = - {1\over 2}$
%(where the alternative quantization would correspond
%to a free fermion field), there is only one quantization.
%Only the pole persists at this value.
%%The values of $k$ at which
%%poles of the Green function occur
%%for both quantizations of a scalar field with
%%$m^2 = -1.3$,
%%as a function of $q$.
%%Note that as $m^2 \to - 5/4$,
%%a line of poles and a zeros of $G$ approach
%%each other.
%%At $m^2 = - 5/4$, there is only one quantization,
%%and the pole persists.
}
\end{center}
\end{figure}

In the intermediate regime $k \in (qe_d, k_{osc})$,
the potential develops a well,
indicating the possible existence of a zero energy bound state.
The locations of these boundstates can easily be determined numerically
and are shown in Fig.~\ref{fig:qkplots}.
Recently, such states have been shown to exist also
for fields near black holes in asymptotically flat space \cite{Hartman:2009qu}.

\vskip-.3in

%$$ \delta_k = \half + \nu_k , ~~~~
%\nu_k = {1\over \sqrt 6} \sqrt{ m^2 + k^2 - q^2/2} $$

%

%\begin{figure}[h!] \begin{center}
%\includegraphics[scale=0.27]{pictures/phase_diagram_half.eps}
%\end{center}
%\end{figure}

\subsection{Summary}

%Low-frequency scaling behavior is determined by dimension of
%operator in IR CFT.

The location of the Fermi surface is determined by short-distance physics
analogous to band structure --
one must find a normalizable solution of $\omega=0$ Dirac equation in full BH.
The low-frequency scaling behavior near the Fermi surface however is universal;
it is determined by near-horizon region and in particular the IR CFT $\sG$.
%\vskip0.2in
Depending on the dimension of the operator in the IR CFT, we find
Fermi liquid behavior (but not Landau) or non-Fermi liquid behavior.
From the bulk point of view,
the quasiparticles decay by falling into the black hole.
The rate at which they fall in is determined by their
effective mass (which by the correspondence determines the exponent $\nu$)
in the near-horizon region of the geometry.

\begin{figure}[h!] \begin{center}
\includegraphics[scale=0.5]{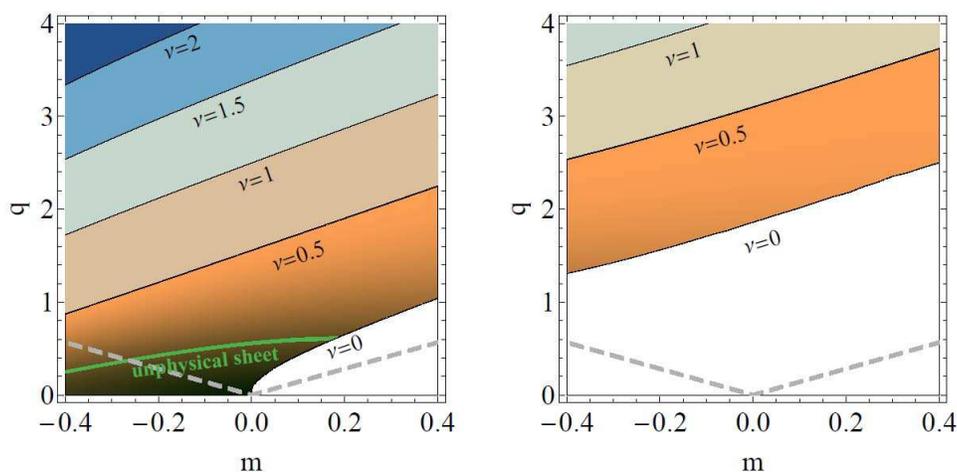}
\caption{\label{fig:phasediagram}
As we vary the mass and charge of the spinor field,
we find the following behavior.
In the white region, there is no Fermi surface.
In the orange region, $\nu< \half$ and we find non-Fermi liquid behavior.
In the remainder of the parameter space, there is a stable quasiparticle.
}
\end{center}
\end{figure}

\section{Correlation functions in the $AdS_2$ IR CFT}
\label{sec:cor}

Here we give a derivation of
the retarded Green's function $\sG(\omega)$
%~\eqref{exbG} and~\eqref{ferG}
for a charged scalar and spinor in AdS$_2$.
We %also discuss their properties and
then describe the finite temperature case.
Then we briefly discuss the generalization to
other IR geometries.

The $AdS_2$ background metric and gauge field given by
   \be \label{aads2}
 {ds^2 } = {R_2^2 \ov \zeta^2 } \(- d \tau^2 +  d \zeta^2 \right), \qquad  A  =  {e_d  \ov  \zeta} d\tau\ .
 \ee
We describe below how the IR CFT correlators relevant to the calculations described above
 can be extracted from the behavior of fields in this background.
One may worry that the nonzero profile of the gauge field breaks the conformal invariance.
In fact, conformal invariance is restored by acting simultaneously with a gauge transformation,
as in \cite{Hartman:2008dq}.

\subsection{Scalar}

We consider the following quadratic scalar action
\be \label{2scaA}
S = -\int  d^{2} x \sqrt{-g} \, \left[g^{ab} (\p_a + i q A_a) \phi^* (\p_b - i q A_b ) \phi  + m^2 \phi^* \phi \right]
\ee
in the background \eqref{aads2}.
In frequency space $\phi (\tau, \zeta) = e^{- i \om \tau} \phi(\om,\zeta)$, the equation of motion for $\phi$ can be written as
 \be \label{Aeep2}
 -\p_\zeta^2 \phi + V (\zeta) \phi =0
 \ee
 with
 \be \label{Aeoep}
 V(\zeta) = {m^2 R_2^2 \ov \zeta^2} - \(\om + {q e_d \ov \zeta} \right)^2 \ .
\ee
Note $\om$ can be scaled away by redefining $\zeta$, reflecting the scaling symmetry of the background solution. Equation~\eqref{Aeep2} can be solved exactly;
the two linearly independent solutions are
\be \phi =  c_{{\rm out}}{\rm W}_{i q e_d, \nu}\left(- { 2 i \omega \zeta } \right)
+c_{{\rm in}}{\rm W}_{-i q e_d, \nu}\left( { 2 i \omega \zeta } \right) ,
\ee
where ${\rm W}_{\lambda,\mu}(z)$ is the Whittaker function, and for scalars
$
\nu \equiv \sqrt{m^2 R_2^2 - q^2 e_d^2 + {1 \ov 4}} ~.
$
%is defined as earlier.
Of these,
the function multiplying $c_{{\rm in}}$,
$ {\rm W}_{i q e_d, \nu}\left(- { 2 i \omega \zeta } \right) \sim e^{ i \omega \zeta} \zeta^{i qe_d} $ is ingoing at the horizon.

Near the boundary of the $AdS_2$ region, the scalar solution behaves as
\be
\phi \buildrel{\zeta \to 0} \over { \approx} A \zeta^{-\nu + \half} + B  \zeta^{\nu+\half} ~~.
\ee
The retarded scalar function of the IR CFT is then\footnote
{Note that we define the scalar Green's function to be $B/A$ without
 the prefactor of $2\nu$ first emphasized in \cite{Freedman:1998tz}.}
 \be \label{AexbG}
 \sG_R (\om) =
 %2\nu
 {B \over A} =
 e^{-i \pi \nu} \frac{ \Gamma (-2\nu ) \Gamma \left(\frac{1}{2}+ \nu-i q e_d\right)}{\Gamma
(2\nu )\Gamma \left(\frac{1}{2}- \nu-i q e_d\right)   } (2\om)^{2 \nu}\  .
 \ee
The advanced function is given by
 \be \label{BAG}
 \sG_A (\om) = e^{i \pi \nu} \frac{ \Gamma (-2\nu ) \Gamma \left(\frac{1}{2}+ \nu + i q e_d\right)}{\Gamma
(2\nu )\Gamma \left(\frac{1}{2}- \nu + i q e_d\right)   } (2\om)^{2 \nu}\ ~~.
 \ee

Now consider a charged scalar field on AdS$_2 \times \RR^{d-1}$.
%It is obvious that
In momentum space the equation of motion reduces to~\eqref{Aeoep} with $m^2$ replaced by
 %introduced in~\eqref{effM}
 \be \label{effM}
 m_k^2 \equiv k^2 {R^2 \ov r_*^2} + m^2, \qquad k^2 = |\vk^2| \ .
 \ee
Thus the retarded function is given by~\eqref{AexbG} with $m^2$ replaced by $m_k^2$.

\subsection{Spinor}

We consider the following quadratic action for a spinor field $\psi$ in the geometry~\eqref{aads2}
  \be \label{ADact}
 S_{\rm spinor} =  \int d^{2} x \sqrt{-g} \, i (\bar \psi \Gamma^a
 %(\p_a - i q A_a)
 D_a \psi  - m \bpsi \psi
 + i \tildem \bpsi \Gamma \psi) ~~;
 \ee
the last term in this action is a parity-violating mass term which in our application will be
related to momentum in $\RR^{d-1}$. We make the following choice of Gamma matrices, chosen to be
compatible with the choice made in \cite{Faulkner:2009wj}, with \eqref{aads2} arising as the near horizon
limit
%\footnote{We denote Pauli matrices by $\hat \sigma$ in order to
%distinguish them from the $\sigma$ near-horizon coordinate.}
\be
\Ga^{\underline \zeta}=  \sigma^3 ,~~~\Ga^{\underline{\tau}} =  i \sigma^1 ,~~~ \Ga =-
\sigma^2~~.
\ee
%\be
%\Ga^r = \left( \begin{array}{cc}
%1 & 0  \\
%0 & -1
%\end{array} \right), \;\;
%\Ga^\mu = \left( \begin{array}{cc}
%0 & \ga^\mu  \\
%\ga^\mu & 0
%\end{array} \right)
%\ee
Then the equations of motion
for $\psi$
can be written as
\be \left( \zeta \partial_\zeta - i   \sigma^3  {\(  \zeta \omega + q e_d \) }
\right) \tilde \Phi
=
{R_2  } \left(   m \sigma^2 + \tildem \sigma^1 \right)
\tilde \Phi~.
\ee
%where
%%$\bar \omega$ is the $\tau$-momentum (related to ordinary frequency by a factor of temperature),
%the gauge field is $A_\tau = e_d\left( {1 \over \sigma} - { 1\over \sigma_0}\right)$,
%and $\bar f \equiv 1 - {\sigma^2 \over \sigma_0^2} $
%is the emblackening factor in the $AdS_2$ black hole metric.
Here
\footnote{There is a similar equation for the component
which computes $G_{2}$ which is obtained by
reversing the sign of the parity-violating mass $\tilde m$.
From the point of view of the $AdS_2$ theory, this is a completely
independent field.
}
\be\tilde \Phi \equiv \left( \begin{matrix} \tilde y \cr \tilde z \end{matrix} \right)
\equiv
 {1\over \sqrt 2} (1 + i \hat \sigma^1)
\left( - g g^{\zeta\zeta}\right)^{-1/4}
\psi
%\left( \begin{matrix}  y_- \cr z_+ \end{matrix} \right)
 .\ee
%\left( g_{\tau\tau} \prod_i g_{ii} \right)^{-1/4}
%\left( \begin{matrix} \psi^y_1 \cr \psi^z_1 \end{matrix} \right) $$

The general solution to this equation is
%at zero temperature ($\zeta_0 \to \infty$) is
%$$
%\tilde \Phi(\sigma) =
% \sigma^{-1/2}
%\left(  c_{{\rm out}}
%\left( \begin{matrix}
%- (  i \tildem  + m ) {\rm W}_{-\half - i qe_d, \nu}(2 i \omega \sigma) \cr
% {\rm W}_{\half - i qe_d, \nu}(2 i \omega \sigma)
%\end{matrix} \right)
%+ c_{{\rm in}} \left( \begin{matrix}
%{\rm W}_{\half + i qe_d, \nu}(-2 i \omega \sigma) \cr
% (  i \tildem  - m ) {\rm W}_{-\half + i qe_d, \nu}(-2 i \omega \sigma)
%\end{matrix} \right)
%\right)
%$$
%or more compactly
\be
\tilde \Phi(\zeta) =
 \zeta^{-1/2}
\left[  c_{{\rm out}}
 {\rm W}_{-{\sigma^3\over 2} - i qe_d, \nu}(2 i \omega \zeta)
 \left( \begin{matrix} \tildem - i m  \cr -1
\end{matrix} \right)
+ c_{{\rm in}} {\rm W}_{{ \sigma^3\over 2} + i qe_d, \nu}(-2 i \omega \zeta)
\left( \begin{matrix}
- 1 \cr \tildem + i m
\end{matrix} \right)
\right]
\ee
%\be
%\tilde \Phi(\zeta) =
% \zeta^{-1/2}
%\left(  c_{{\rm out}}
% {\rm W}_{{\hat \sigma^3\over 2} - i qe_d, \nu}(2 i \omega \zeta)
% \left( \begin{matrix} 1  \cr \tildem + i m
%\end{matrix} \right)
%+ c_{{\rm in}} {\rm W}_{-{\hat \sigma^3\over 2} + i qe_d, \nu}(-2 i \omega \zeta)
%\left( \begin{matrix}
%\tildem - i m  \cr 1
%\end{matrix} \right)
%\right)
%\ee
%\be
%\tilde \Phi(\zeta) =
% \zeta^{-1/2}
%\left(  c_{{\rm out}}
% {\rm W}_{{\hat \sigma^3\over 2} - i qe_d, \nu}(2 i \omega \zeta)
% \left( \begin{matrix} \tildem + i m   \cr 1
%\end{matrix} \right)
%+ c_{{\rm in}} {\rm W}_{-{\hat \sigma^3\over 2} + i qe_d, \nu}(-2 i \omega \zeta)
%\left( \begin{matrix}
%1 \cr \tildem - i m
%\end{matrix} \right)
%\right)
%\ee
%The ingoing solution to this equation at zero temperature
%is
%$$
%\tilde \Phi(\sigma) =
% \sigma^{-1/2}
%\left( \begin{matrix}
% {\rm W}_{\half - i qe_d, \nu}(2 i \omega \sigma) \cr
%(  i \tildem  - m )  {\rm W}_{-\half - i qe_d, \nu}(2 i \omega \sigma)
%\end{matrix} \right)
%$$
where again $ {\rm W}$ is a Whittaker function,
and where, for spinors, $ \nu = \sqrt{m^2 R_2^2 - q^2 e_d^2} $.
The notation $\sigma^3$ in the index of the Whittaker function
indicates $\pm 1 $ when acting on the top/bottom component of the spinor.

%\sigma^{-\half + m\ell_2} W_{\half + i qe_d, ~2 \nu}(- 2 i \omega \sigma)
%+ (m-i \tildem) \ell_2 W_{- \half + i qe_d, ~2 \nu}(- 2 i \omega \sigma). $$
%The function whose coefficient is $c_{{\rm in}}$ is ingoing.

\def\UUU{\aleph}
The field which matches to the spinor field in the outer region
%in the basis
%%\eqref{realbasis}
is
$\Phi = {1\over \sqrt2} (  1 - i \sigma^1) \tilde \Phi$.
%The asymptotic behavior of $\Phi$ near the boundary of the $AdS_2$ region is
%as in equation \eqref{eqn:nearUV}
%%\eqref{spinormatch} and \eqref{NHspinorasymptotics}.
%$$ \Phi \buildrel{\zeta \to 0} \over { \simeq}
%\UUU
%\left( \begin{matrix}
% \zeta^{-\nu} \cr
%\zeta^{\nu}
%\end{matrix} \right)
%~;
%$$
%in terms of the asymptotics of the Whittaker function
%$$W_{\pm {1\over 2} + i q e_d , \nu}(-2 i \omega \zeta)
%\buildrel{\zeta \to 0} \over { \simeq} A_{\pm } \zeta^{-\nu} + B_{\pm} \zeta^{\nu} $$
%the matrix $\UUU$ takes the form
%$$
%\UUU =
%\left( \begin{matrix}
%- i A_{+} + (m - i\tildem ) A_{-} & -i B_{+} + (m- i \tildem ) B_{-}
%\cr
%A_{+} - i  (m - i \tildem) A_{-} & B_{+} -i  (m - i \tildem ) B_{-}
%\end{matrix} \right)
%$$
%The fact that one gets the same answer for $\sG$
%from either component is a consistency check on the matching
%procedure.
%SNIP
Near the boundary of $AdS_2$, $\zeta\to 0$, the $AdS_2$ Dirac equation in this basis becomes
 \be
 \zeta \partial_\zeta \Phi =  U \Phi , \quad U =  \left( \begin{matrix}
m R_2 & \tilde m R_2 - q e_d
\cr
\tilde m R_2 + q e_d  & -m R_2
\end{matrix} \right) \ .
 \ee
%Thus as $\ze \to 0$,
The asymptotic behavior of $\Phi$ near the boundary of the $AdS_2$ region is
as in equation \eqref{eqn:nearUV}:
%$\Phi$ can be written as
 \be
 \Phi = A \, v_- \zeta^{- \nu} \(1 + O(\zeta) \)+
 B \, v_+ \zeta^{\nu} \(1 + O(\zeta) \)
 \ee
where $v_\pm$ are real eigenvectors of $U$ with eigenvalues $\pm \nu$.
The relative normalization of $v_+$ and $v_-$ is a convention, which affects the normalization of the $AdS_2$ Green's functions.
We will choose $v_\pm$ to be given by
 \be \label{eignV}
v_\pm =   \left( \begin{matrix}
m R_2 \pm \nu
\cr
\tilde m R_2 + q e_d
\end{matrix} \right) \ .
\ee
With the normalization convention in \eqref{eignV},
the bottom components of $v_\pm$ are equal and
so $A, B$ can be extracted from the asymptotics of $z$.
The relative normalization of the eigenspinors
in \eqref{eignV} affects the answer for the $AdS_2$ Green's function;
$v_\pm \to \lambda_\pm v_{\pm}$ takes $\sG_R \to \frac{\lambda_+}{\lambda_-} \sG_R$.
This rescaling does not, however, affect the full Green's function $G_R$
computed by the matching procedure.

One then finds that the retarded Green's function
of the operator coupling to $y$ (the upper component of $\Phi$)
can be written as
%({\bf check prefactor})
 \be \label{AeFG}
 \sG_R (\om) = e^{- i \pi \nu} \frac{\Gamma (-2 \nu ) \, \Gamma \left(1+\nu -i q e_d \right)}{\Gamma (2 \nu )\,   \Gamma \left(1-\nu -i q e_d\right)}
% \frac{m R_2 - i q e_d - \nu}{ m R_2 - i q e_d +  \nu} \( 2\om\)^{2 \nu}.
\cdot \frac{ \le(m - i \tildem \ri) R_2- i q e_d - \nu}
{  \le(m - i \tildem \ri) R_2 - i q e_d +  \nu} \( 2\om\)^{2 \nu}.
 \ee
The advanced function is given by
 \be \label{AAG}
 \sG_A (\om) = e^{ i \pi \nu} \frac{\Gamma (-2 \nu ) \, \Gamma \left(1+\nu + i q e_d \right)}{\Gamma (2 \nu )\,   \Gamma \left(1-\nu +i q e_d\right)}
 \cdot
 \frac{ \le(m + i \tildem \ri) R_2 +i q e_d - \nu}
 {  \le(m+ i \tildem \ri) R_2 + i q e_d +  \nu} (2\om)^{2 \nu}
 \ee

%{\tt[This needs to be rechecked.]}
%Next we wish to determine the phase of $\sG$ outside
%the oscillatory region, and its magnitude inside the oscillatory region.
%Note the identity
% \be \label{oo}
% {\sG_R (\om) \ov \sG_A (\om)} = - e^{-2 \pi i \nu} {\sin \pi (\nu + iq e_d) \ov \sin \pi (\nu - i q e_d )} = {e^{-2 \pi i \nu} - e^{-2 \pi q e_d} \ov e^{2 \pi i \nu} - e^{-2 \pi q e_d}} \ .
% \ee
%Similarly writing $\sG_R$ in the form~\eqref{oepB},  from~\eqref{oo} we find that for real
%$\nu > 0$
%  \be
% \ga = \arg\left( \Gamma(-2\nu) \le(e^{-2 \pi i \nu} - e^{- 2 \pi q e_d} \ri)\right)
% \ee
% which implies that $ \ga < -2 \pi \nu $. For $\ga = - i\lam$ pure imaginary we
% find from~\eqref{anid}
% \be
% c^2 = {e^{-2 \pi \lam} - e^{-2 \pi q e_d} \ov e^{2 \pi \lam} - e^{-2 \pi q e_d}}
% < e^{- 4 \pi \lam} \ .
% \ee

Now consider a charged fermion field on AdS$_2 \times \RR^{d-1}$. Taking only $k_1 = k\neq 0$ dimensional reduction to~\eqref{Aeoep} we find the action given by
\eqref{ADact}
with the identification $\tildem = - k {r_\star \over R}(-1)^\alpha$.

%\clearpage

\subsection{Finite temperature generalization}
\label{sec:ads2T}

\def\ze{\zeta}
The above discussion can be generalized to finite temperature.
In this case, the near horizon region is a (charged) black hole in $AdS_2$ (times space):
 \be \label{ads2T}
 ds^2 =  {R^2_2 \ov \ze^2} \le( - \le(1- {\ze^2 \ov \ze_0^2} \ri)  d\tau^2 +
 {d \ze^2 \ov 1- {\ze^2 \ov \ze_0^2}} \ri)
 + {r_*^2 \ov R^2} d \vec x^2
% \ee
% with
%\be
~~~~
A =  {e_d  \ov \ze}  \le(1-{\ze \ov \ze_0} \ri) d\tau
\ee
and a temperature (with respect to $\tau$)
$
 T ={1 \ov 2 \pi \ze_0} \ .
$
Locally, this black hole in $AdS_2$ is related by a coordinate transformation
(combined with a gauge transformation) to
the vacuum $AdS_2$, that is, it can be obtained by some global identifications.
This fact can be used to infer the finite-temperature correlators from their
zero-temperature limit presented above; they can also be determined directly
as we do next.

The equation of motion for a minimally coupled scalar is
% {\footnotesize
\be
  \nonumber
   \p_\zeta^2\phi  +  \frac{ 2 \zeta   }{\zeta^2-\zeta_0^2}\p_\zeta\phi  +
   \left(-\frac{  m^2 R_2^2}{\zeta^2-{\zeta^4 \ov \zeta_0^2}} + \frac{\left[\omega + q e_d \left({1\ov \zeta}- {1\ov \zeta_0}\right)\right]^2}{
1-{2\zeta^2\ov \zeta_0^2}+{\zeta^4\ov \zeta_0^4}} \right)\phi = 0~.
\ee
%}
Two linearly independent solutions are given by
%{\footnotesize
\be
 \nonumber \phi(\zeta) \sim \, \left({1\ov \zeta}- {1\ov \zeta_0} \right)^{-{1 \over 2} \mp \nu} \left({\zeta_0 + \zeta \ov \zeta_0 - \zeta}\right)^{ { i
\omega \zeta_0 \ov 2}- i q e_d }
% \times  \qquad \qquad \qquad
%\\
% \nonumber
{}_2F_1\left({1 \over 2} \pm \nu
+ {i \omega \zeta_0 }- i q e_d , {1 \over 2} \pm \nu - i q e_d  , \  1 \pm 2 \nu; \ {2\zeta
\ov \zeta - \zeta_0} \right)
\ee
%}
where
$
  \nu =  \sqrt{{1 \over 4  }+m^2 R_2^2 - q^2 e_d^2 }
$
and the top sign gives the ingoing solution.

Using $T = {1 \ov 2\pi \zeta_0}$, the scalar retarded Green's function has the following form
%{\footnotesize
 \be
   \nonumber
 \sG_R (\om) = (4 \pi T)^{2 \nu} \frac{ \Gamma (-2\nu ) \Gamma \left(\frac{1}{2} +\nu-\frac{i \omega }{2 \pi T }+
i q e_d\right)\Gamma \left(\frac{1}{2}+ \nu-i q e_d\right)}{\Gamma
(2\nu )\Gamma \left(\frac{1}{2}-\nu -\frac{i \omega }{2 \pi T }+i q e_d
\right)\Gamma \left(\frac{1}{2}- \nu-i q e_d\right)   } ~.
 \ee %}

The spinor equation of motion at finite temperature is
\be \left(  \partial_\zeta - i  \hat \sigma^3  { \omega + q A_\tau \over \bar f}
\right) \tilde \Phi
=
{R_2 \over   \zeta  \sqrt {\bar f} } \left( m \hat \sigma^2 + \tildem \hat \sigma^1 \right)
\tilde \Phi~
\ee
%where
%%$\bar \omega$ is the $\tau$-momentum (related to ordinary frequency by a factor of temperature),
%the gauge field is $A_\tau = e_d\left( {1 \over \sigma} - { 1\over \sigma_0}\right)$,
%and
where $\bar f \equiv 1 - {\zeta^2 \over \zeta_0^2} $
is the emblackening factor in the $AdS_2$ black hole metric in \eqref{ads2T}.
It can similarly be solved in terms of hypergeometric functions.
The general solution for the upper component $\tilde y$ of $\tilde \Phi$ is
%{\footnotesize
\bea
\tilde y(\zeta) &=& \( \frac{\zeta + \zeta_0 }{\zeta} \)^{\half + i \( qe_d + \frac{\zeta_0 \omega}{2} \)}
\cr
&&
\left[
c_{\text{in}} \( - 1 + \frac{\zeta_0}{\zeta} \)^{-\frac{i \zeta_0 \omega}{2} }
{}_2 F_1 ( \half + i q e_d - \nu - i \zeta_0 \omega,
\half + i q e_d + \nu - i \zeta_0 \omega, \half - i \zeta_0 \omega, \frac{\zeta - \zeta_0 }{2 \zeta} )
\right.
\cr
&+&
\left.
c_{\text{out}}  \( - 1 + \frac{\zeta_0}{\zeta} \)^{ \frac{i \zeta_0 \omega}{2} + \frac{1}{2}}
{}_2 F_1 ( 1 + i q e_d - \nu ,
1 + i q e_d + \nu , \frac{3}{2} + i \zeta_0 \omega, \frac{\zeta - \zeta_0 }{2 \zeta} )
\right].
%&&
%\left[
%c_{\text{in}} \( - 1 + \frac{\zeta_0}{\zeta} \)^{- \frac{i \zeta_0 \omega}{2} }
%{}_2 F_1 ( \half + i q e_d - \nu - i \zeta_0 \omega,
%\half + i q e_d + \nu - i \zeta_0 \omega, \half - i \zeta_0 \omega, \frac{\zeta - \zeta_0 }{2 \zeta} )
%\right.
%\cr
%&+&
%\left.
%c_{\text{out}}  \( - 1 + \frac{\zeta_0}{\zeta} \)^{ \frac{i \zeta_0 \omega}{2} + \frac{i \zeta_0 \omega}{2}}
%{}_2 F_1 ( 1 + i q e_d - \nu ,
%1 + i q e_d + \nu , \frac{3}{2} + i \zeta_0 \omega, \frac{\zeta - \zeta_0 }{2 \zeta} )
%\right].
\eea
%}
The
retarded function is then
\be
%\nonumber
 \sG_R (\omega)= (4 \pi T)^{2 \nu} \frac{\( m - i \tilde m\) R_2+i q e_d + \nu}{ \( m - i \tilde m\) R_2+ i q e_d -  \nu}
 %\times \qquad  \qquad \qquad
 %\\ \nonumber
% \times
\cdot
%\quad
  {\Gamma (-2 \nu ) \Gamma (\half+\nu -\frac{i \omega }{2 \pi T }+i q e_d )\,  \Gamma \left(1+\nu -i
q e_d \right)\ov \Gamma (2 \nu )\Gamma \left(\frac{1}{2}-\nu -\frac{i \omega }{2 \pi T}+i q e_d
\right)\, \Gamma \left(1-\nu -i q e_d \right)}
\ee
%\bea
%\nonumber
% \sG_R (\omega)= (2 \pi T)^{2 \nu} \frac{m R_2- i q e_d - \nu}{ m R_2- i q e_d +  \nu} \times \qquad  \qquad \qquad \\ \nonumber
% \times
%  {\Gamma (-2 \nu ) \Gamma (\half+\nu -\frac{i \omega }{2 \pi T }+i q e_d )\,  \Gamma \left(1+\nu -i
%q e_d \right)\ov \Gamma (2 \nu )\Gamma \left(\frac{1}{2}-\nu -\frac{i \omega }{2 \pi T}+i q e_d
%\right)\, \Gamma \left(1-\nu -i q e_d \right)}
%\eea

Note that the branch point at $\om =0$ of zero temperature now disappears and
the branch cut is replaced at finite temperature by a line of poles parallel to the next imaginary axis. In the zero temperature limit the pole line becomes a branch cut.
Similar phenomena have been observed previously \cite{Festuccia:2005pi}.

\subsection{Finite temperature correlators and conformal invariance}

In this subsection we describe how conformal invariance of the IR CFT actually completely fixes the frequency dependence of the finite temperature correlator derived above. The key point here, described in \cite{Spradlin:1999bn}, is that the $AdS_2$ black hole metric \eqref{ads2T} is actually related to the $T = 0$ $AdS_2$ metric via a coordinate transformation that acts as a conformal transformation on the boundary of $AdS_2$.  To see this explicitly, consider the following coordinate transformation to a new set of coordinates $(\sig, t)$:
\be
\tau \pm \xi_0 \tanh^{-1}\le(\frac{\xi}{\xi_0}\ri) = \xi_0 \log \le(\frac{t \pm \sig}{\xi_0}\ri). \label{coord1}
\ee
Working out the metric \eqref{ads2T} in these new coordinates we find eventually:
\be
ds^2 = R_2^2 \frac{-dt^2 + d\sig^2}{\sig^2} \label{zeroTmetric}
\ee
This is just the $T = 0$ $AdS_2$ metric, as claimed above. To better understand what happened consider the effect of the coordinate transformation \eqref{coord1} at the AdS$_2$ boundary $\sig = \xi = 0$
\be
t = \frac{1}{2\pi T}\exp(2\pi T \tau), \label{conftrans}
\ee
which is exactly the transformation that generates Rindler space in the $\tau$ coordinate from the vacuum in the $t$ coordinate. This is simply the statement that a coordinate choice that defines a black hole in the $AdS_2$ bulk is equivalent to a coordinate choice that puts the field theory at a finite temperature. Because of conformal invariance this coordinate change is actually a symmetry operation in the IR CFT.

We note however that the gauge field adds a new subtlety; in particular, going through the same procedure with the zero temperature gauge field \eqref{aads2} does \emph{not} give us the finite temperature gauge field \eqref{ads2T}. Let us go through the same steps as previously, starting this time with the zero temperature configuration appropriate to the metric \eqref{zeroTmetric}:
\be
A = \frac{e_d}{\sigma} dt
\ee
writing this in terms of the finite-temperature coordinates we obtain after some algebra
\be
A = \frac{e_d}{\xi} d\tau + e_d d\le(\xi_0 \tanh^{-1}\le(\frac{\xi}{\xi_0}\ri)\ri)
\ee
Compare this to the gauge field configuration in \eqref{ads2T}. It is not the same, but the difference is pure gauge; indeed the gauge field $\bar{A}$ defined by
\be
\bar{A} = A + d\Lambda \qquad \Lambda = -2\pi T e_d \tau - e_d \le(\xi_0 \tanh^{-1}\le(\frac{\xi}{\xi_0}\ri)\ri)
\label{gaugetrans}
\ee
\emph{is} exactly the gauge field appearing in the finite-temperature solution \eqref{ads2T}. Thus we have shown that the charged $AdS_2$ finite temperature black hole is precisely the same as a coordinate transform {\it plus} a gauge transform of the vacuum $AdS_2$. Note that this gauge transform does not vanish at the $AdS_2$ boundary $\xi = 0$; the $d\tau$ part remains nonzero and corresponds to putting the IR CFT field theory at an (extra) constant value of $A_\tau$.

Now the transformation of the IR CFT correlators under each of these manipulations is known; thus it should be possible to determine the finite temperature correlators from the $T = 0$ conformal results. For the conformal transformation \eqref{conftrans} we use the usual CFT tranfsormation law
\be
\langle \sO^{\dagger}(\tau) \sO(\tau') \rangle = \le(\frac{dt}{d\tau}\ri)^{\delta}\le(\frac{dt'}{d\tau'}\ri)^{\delta}\langle \sO^{\dagger}(t) \sO(t') \rangle,
\ee
where $\delta = \frac{1}{2} + \nu$ is the dimension of the IR CFT operator.

The operation \eqref{gaugetrans} is different and corresponds to turning on a source for the boundary gauge field that is pure gauge: generalizing slightly, in higher dimensions it would correspond to $A_{\mu} = \partial_{\mu}\Lam$, with $\mu$ running over all field theory directions (in our case, only $t$). Thus with the insertion of this source we are now computing the field theory correlator
\be
\langle \sO^{\dagger}(x) \sO(x') \rangle_{\Lam} = \left\langle \exp \left(-i\int dy\;\Lam(y) \p_\mu j^{\mu}(y)\right) \sO^{\dagger}(x) \sO(x') \right\rangle
\ee
$\Lam(x')$ has an effect on this correlator only because of the contact terms in correlation functions of the divergence of a current with a charged operator, e.g. $i\langle \p_{\mu}j^{\mu}(y) \sO(x) \rangle = i q\langle \sO(x)\rangle \delta(x-y)$. In our case we find
\be
\langle \sO^{\dagger}(x) \sO(x') \rangle_{\Lam} = \exp\le(i q\le(\Lam(x) - \Lam(x')\ri)\ri)\langle\sO^{\dagger}(x) \sO(x')\rangle_{\Lam = 0}
\ee
Putting these ingredients together and using the fact that in the $(t,\sigma)$ coordinate system we have
just $\langle \sO^{\dagger}(t) \sO(t') \rangle \sim (t-t')^{-2\delta}$ we find

\be
\langle \sO^{\dagger}(\tau) \sO(\tau') \rangle \sim \left(\frac{\pi T}{\sinh(\pi T(\tau-\tau'))}\right)^{2\delta}\exp(-2\pi i T e_d q (\tau - \tau'))
\ee
Here the first factor is simply the usual expression for a finite temperature correlator in the chiral half of a 2d CFT in a thermal ensemble, where the role of the coordinate of the ``chiral half'' is being played by the time coordinate. The extra factor is a new ingredient arising from the nontrivial sourcing of the gauge field. Upon Fourier transformation
%\footnote{This is a very difficult Fourier transform.} 
we should find the usual expression from 2d CFT at finite temperature (see e.g. \cite{Iqbal:2009fd, Birmingham:2001pj}), except with a shift in the frequency from the gauge field contribution:
\be
\sG(\om) \sim (2\pi T)^{2\delta -1}\frac{\Gamma\left(\delta - \frac{i}{2\pi T}(\om - 2\pi q e_d T)\right)}{\Gamma\left(1 - \delta - \frac{i}{2\pi T}(\om - 2\pi q e_d T)\right)}
\ee
This frequency dependence agrees with that of the correlator derived directly from the bulk wave equation above.

\section{Finite temperature}

\begin{figure}[h]
\begin{center}
  \includegraphics[totalheight=3.5cm,origin=c,angle=0]{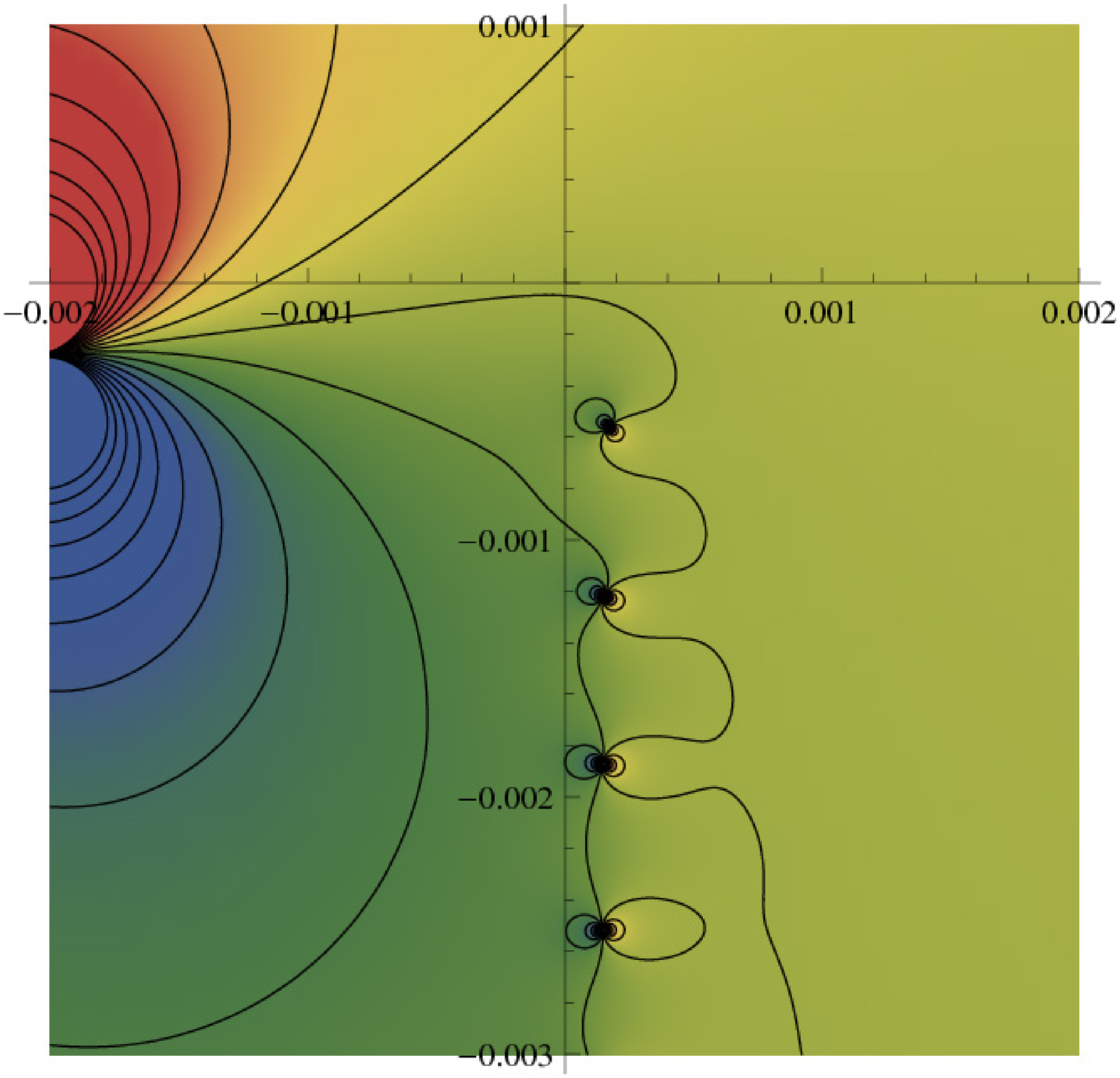}
  \includegraphics[totalheight=3.5cm,origin=c,angle=0]{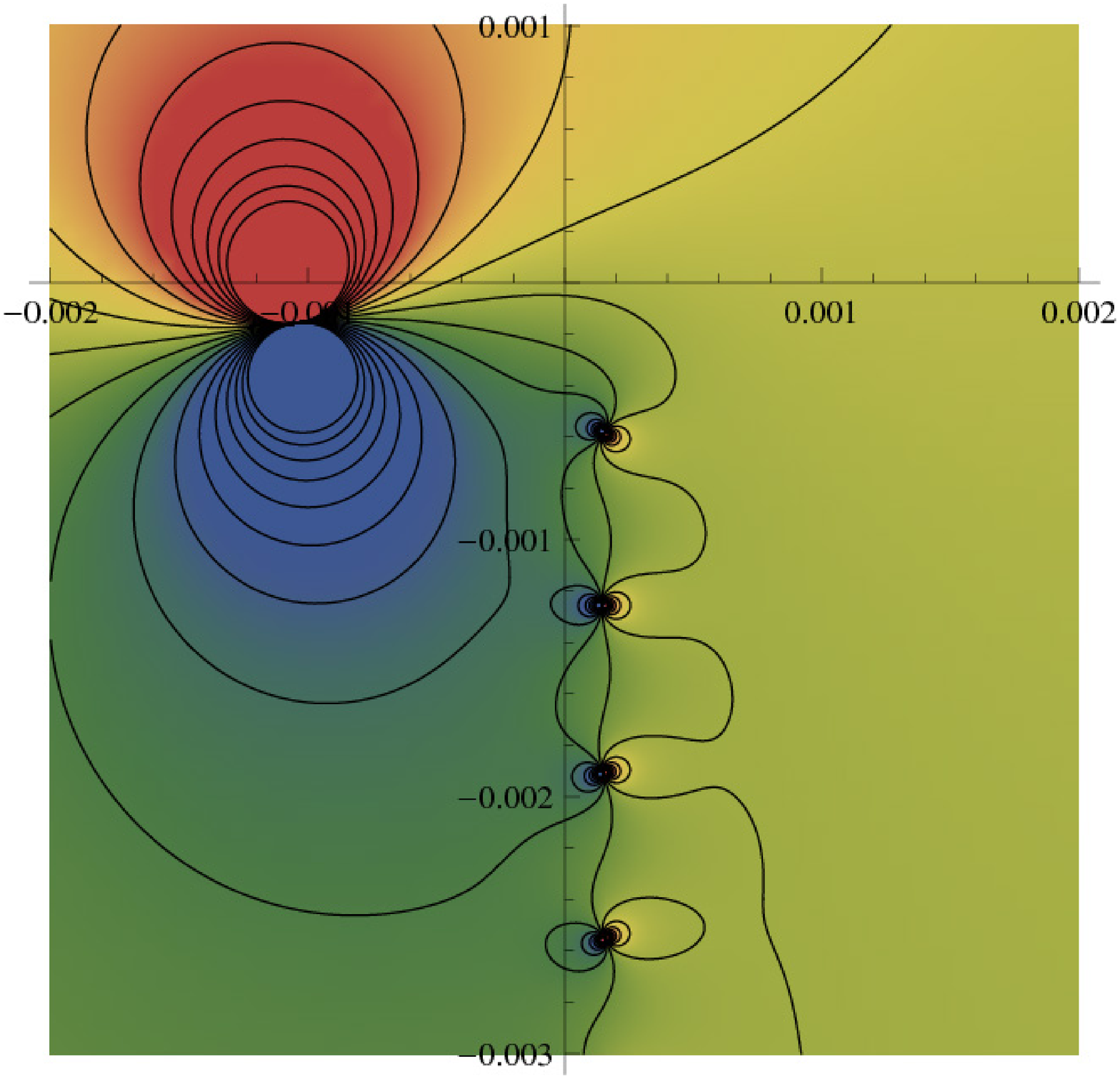}
 \includegraphics[totalheight=3.5cm,origin=c,angle=0]{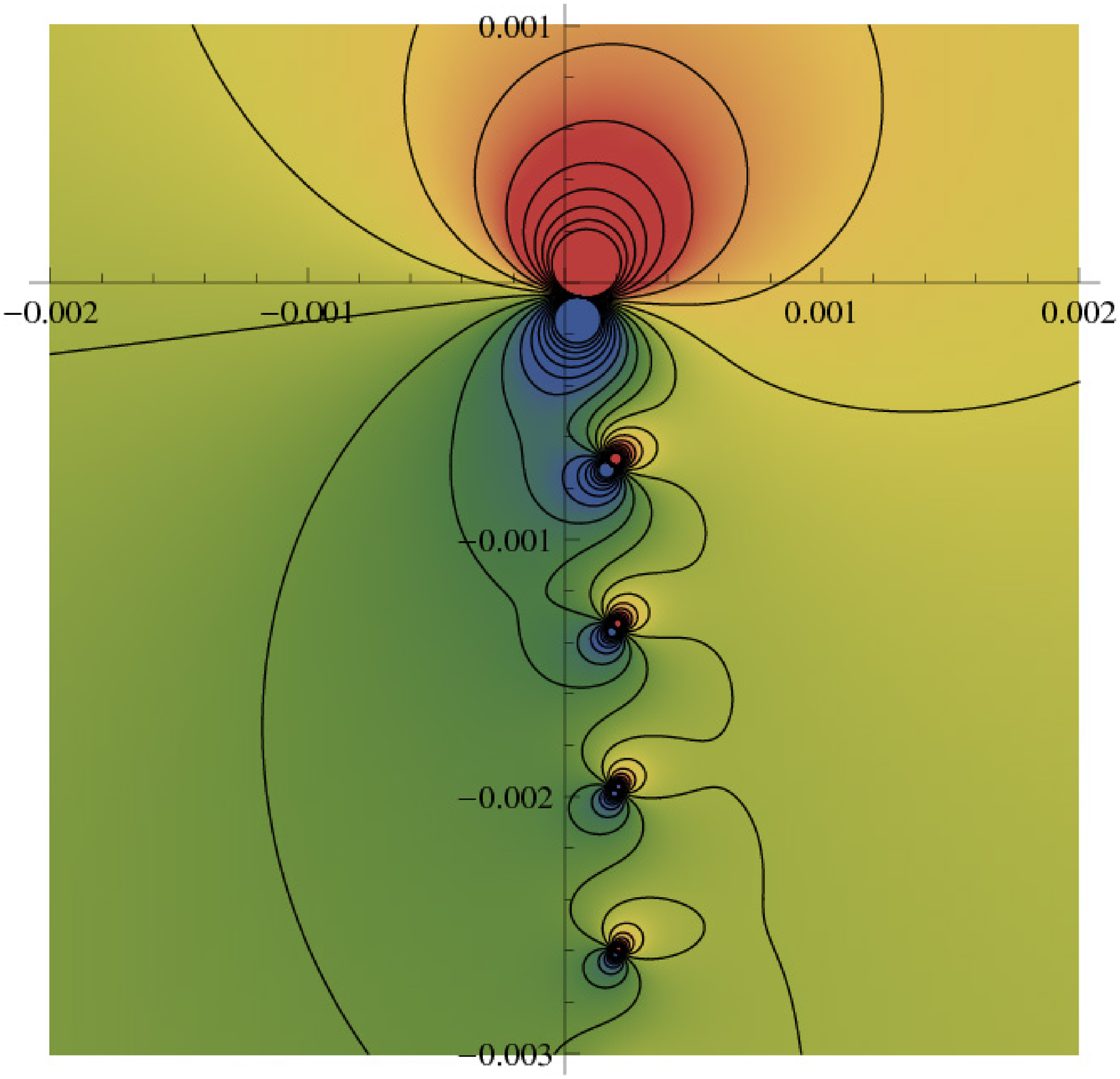}\\
 \includegraphics[totalheight=3.5cm,origin=c,angle=0]{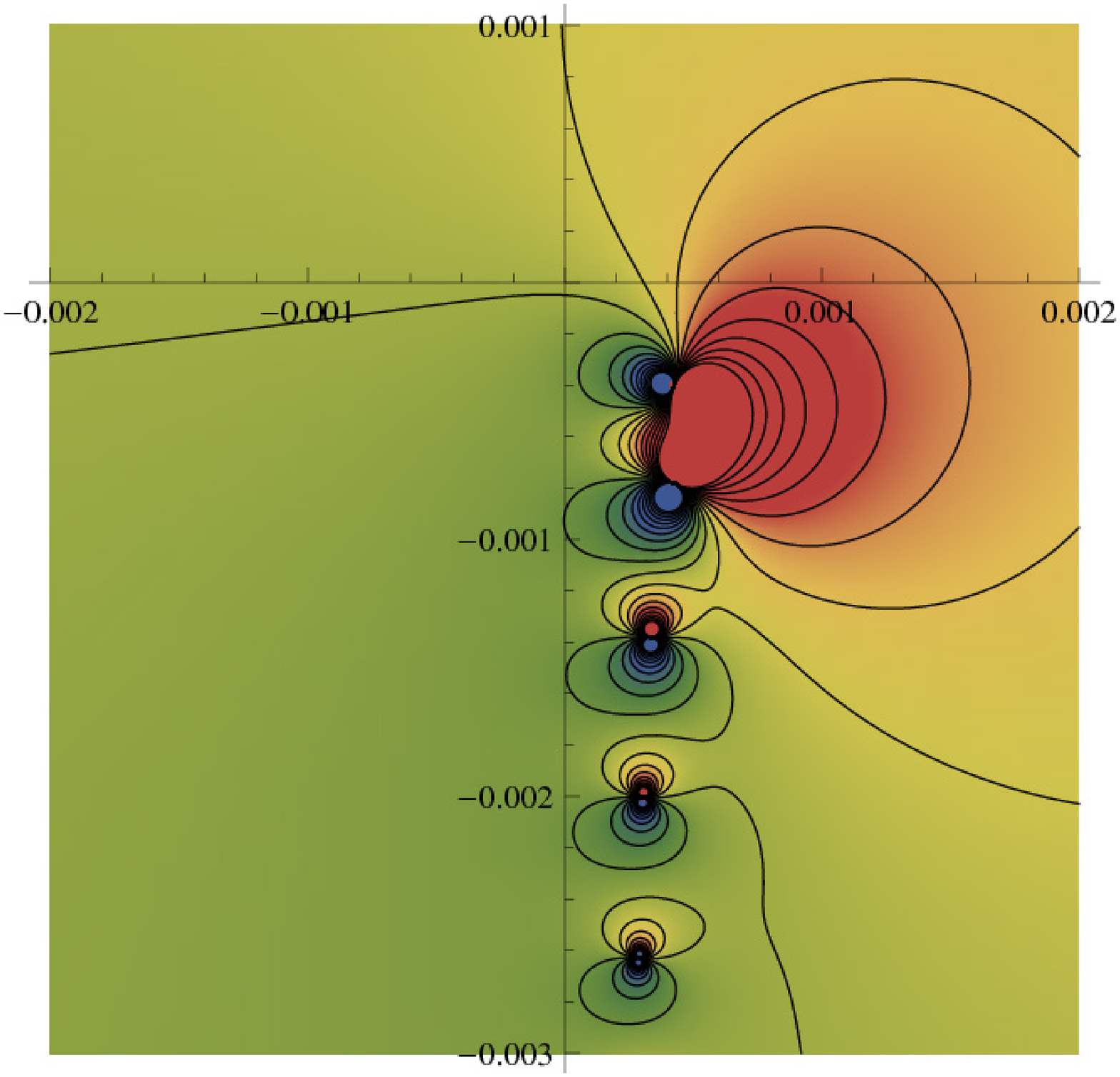}
 \includegraphics[totalheight=3.5cm,origin=c,angle=0]{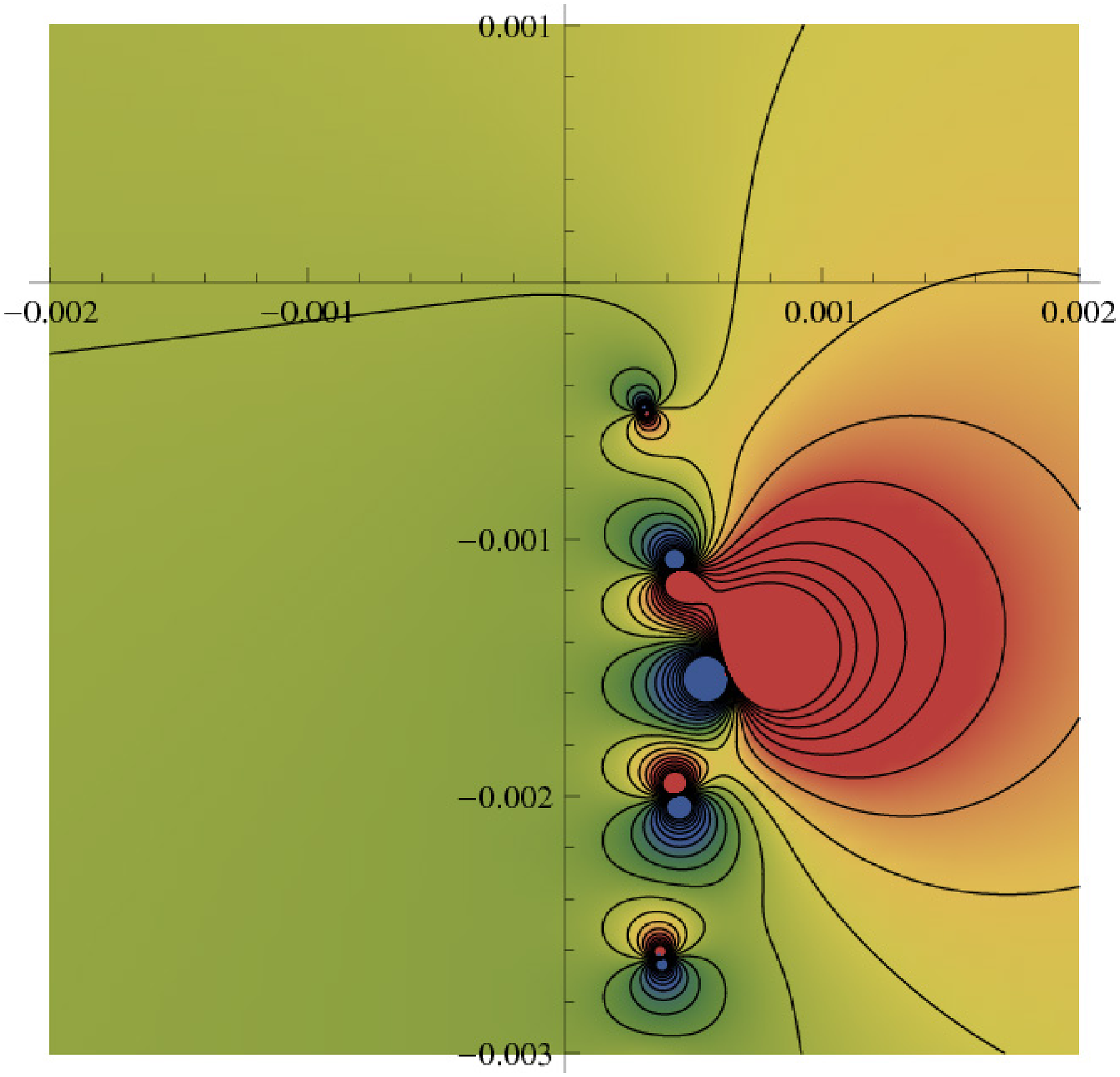}
 \includegraphics[totalheight=3.5cm,origin=c,angle=0]{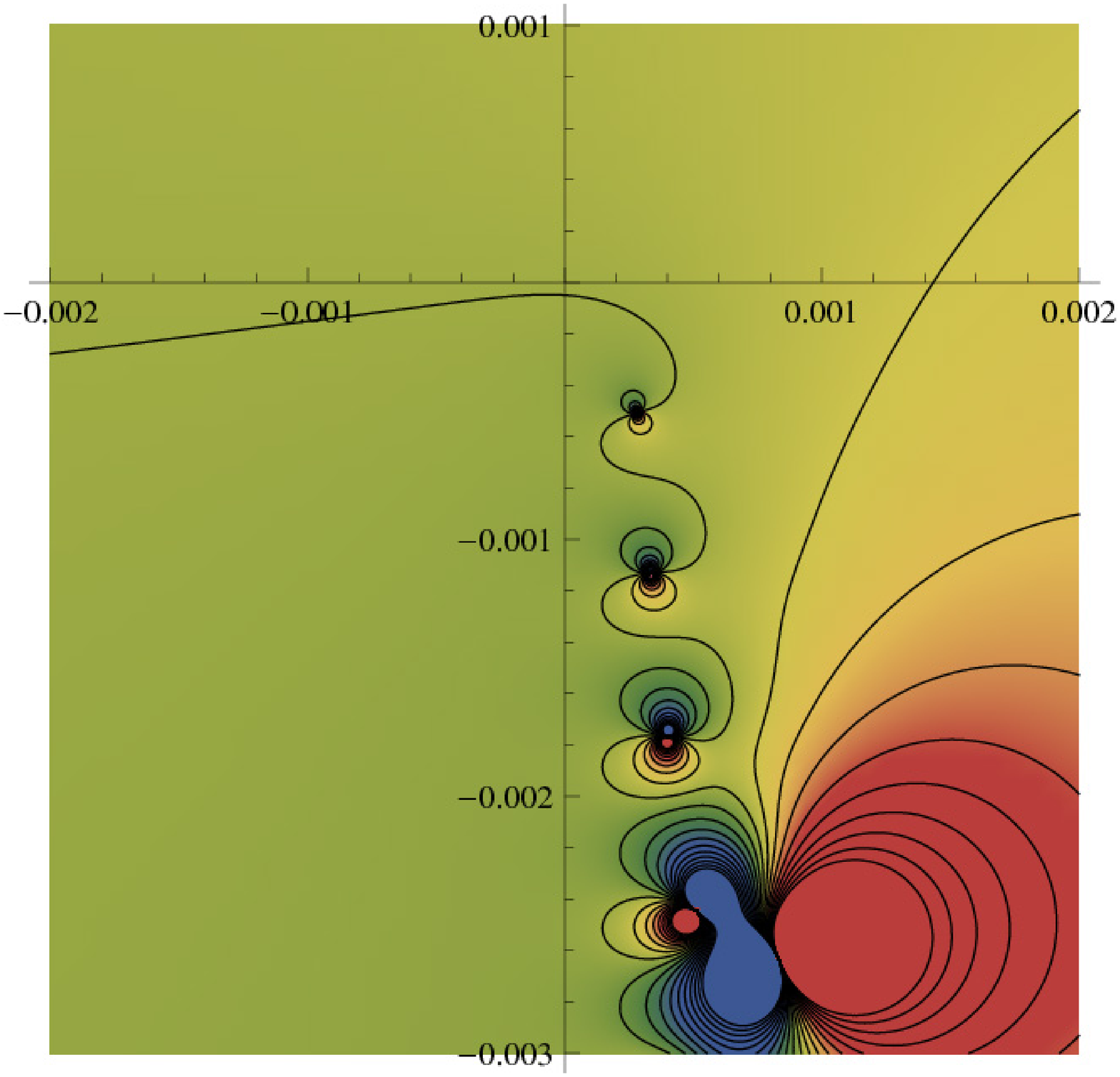}
 \caption{\label{complex_omega}
Density plots of $\Im G_R$ in the complex frequency plane for a sequence of values of $k$
near $k_F$, at finite temperature.
%The parameters are $m=0, q=1, T = ?$, and this is the $\alpha = 1$ component [?].  For these values $k_F = 1.080123$;
%in the plot $k$ runs from $.75$ to $1.5$.
The parameters are $m=0, \, q=1.23, \, T = 10^{-4}$ and this is the $\alpha = 2$ component;
at $T=0$ for these parameters, the Fermi surface lies at
$k_F = 1.20554$. In the plots, $k_\perp = -0.01  \ldots  0.03$.
} % $k_F = 1.2055366568923817282785199377$
\end{center}
\end{figure}
%\vskip-.2in

%\begin{figure}[h]
%\begin{center}
%  \includegraphics[totalheight=2.91cm,origin=c,angle=0]{pictures/complex_omega_with_path_colored2.eps}
%%\caption{\label{complex_omega}
%%}
%\end{center}
%\end{figure}
%\vskip-.2in
%[FIX PARAMETER VALUES.]
%The complex omega plane for $T=4.13 \times 10^{-4}$:
%%the quasi-particle pole is a finite distance {\it below} the real $\om$-axis. \\
%\textcolor{darkblue}{dashed line:}
%\bluecom{trajectory of the pole between $k=0.87$(left)$\ldots 0.93$(right). }
%\greencom{${\rm min}_k\(\Im \omega_c\) \simeq T$ (up to 1\% accuracy). }
%\textcolor{darkblue}{In background:}
%\greencom{density plot for $\Im G_{22}(\omega)$ at $k = 0.90$}
%\greencom{where the corresponding pole is closest to the real axis.}
%
%\begin{figure}[h]
%\begin{center}
%  \includegraphics[totalheight=2.9cm,origin=c,angle=0]{pictures/complex_omega_with_path_colored.eps}
%  \caption{\label{complex_omega}The complex omega plane. The density plot shows $-\Im G_{22}(\omega)$ for $k = 0.90$, $q=1$
%  and $T=4.13 \times 10^{-4}$. The dashed line indicates the trajectory of the pole between $k=0.87$(left)$\ldots 0.93$(right). The closest
%  distance to the real axis is equal to the temperature (up to 1\% accuracy).}
%\end{center}
%\end{figure}

At small nonzero temperature, the near-horizon geometry is a black hole in $AdS_2$.
\begin{figure}[h] \begin{center}
\includegraphics[scale=0.30]{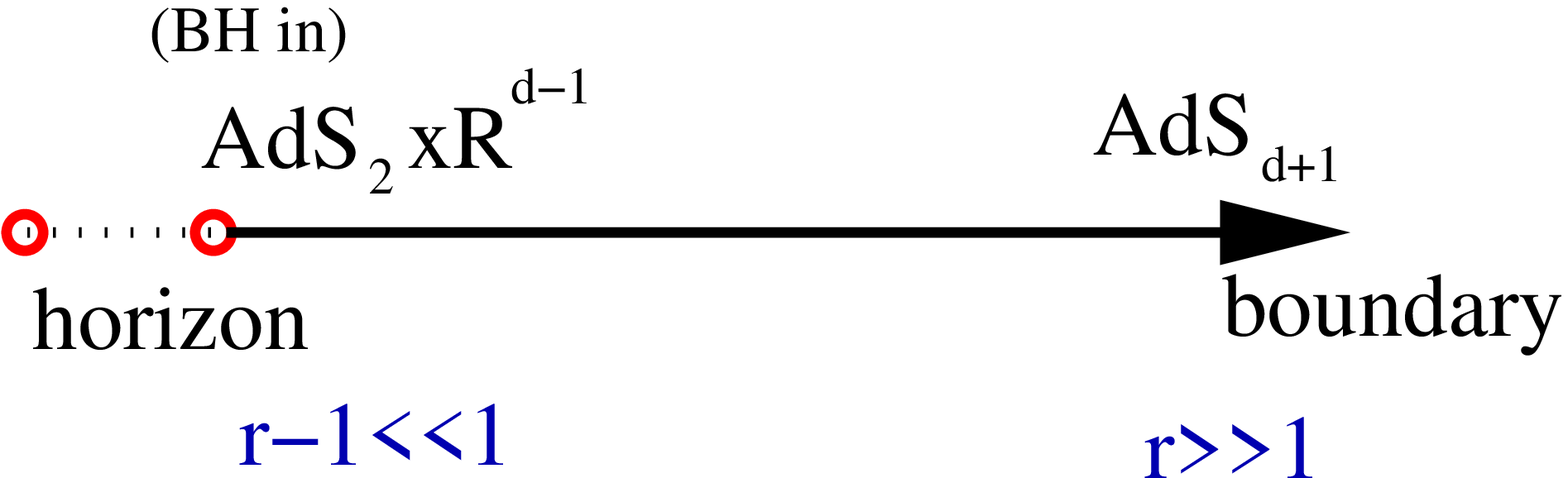}
\end{center}
\end{figure}
%\vskip-.2in
Using the results for $\sG(\omega, T)$ from Section~5\ref{sec:ads2T} for the
Green's functions resulting from this IR geometry,
the fermion self energy becomes
%One finds for a scalar
% \bwt
% \be
% \sG_R (\om) = (2 \pi T)^{2 \nu} \frac{ \Gamma (-2\nu ) \Gamma \left(\frac{1}{2} +\nu-\frac{i \omega }{2 \pi T }+
%i q e_d\right)\Gamma \left(\frac{1}{2}+ \nu-i q e_d\right)}{\Gamma
%(2\nu )\Gamma \left(\frac{1}{2}-\nu -\frac{i \omega }{2 \pi T }+i q e_d
%\right)\Gamma \left(\frac{1}{2}- \nu-i q e_d\right)   }
% \ee
% \ewt
%and for a spinor
%\bwt
%$$
% \sG_R (\omega)= (2 \pi T)^{2 \nu} {\Gamma (-2 \nu )  \ov \Gamma (2 \nu )}
%  {\Gamma (\half+\nu -\frac{i \omega }{2 \pi T }+i q e_d )\,  \Gamma \left(1+\nu -i q e_d \right)\ov \Gamma \left(\frac{1}{2}-\nu -\frac{i \omega }{2 \pi T}+i
%q e_d \right)\, \Gamma \left(1-\nu -i q e_d \right)}
%\frac{(m + i \tilde m) R_2-
%i q e_d - \nu}{ (m + i \tilde m) R_2- i q e_d +  \nu}
%$$
%\ewt
%\vskip.1in
%\greent{$\omega^{2\nu}$ is the $T\to 0$ limit of }
\be
\label{eq:sigT}
 \Sigma(\omega, T) = T^{2\nu} g(\omega/T)= (4 \pi T)^{2 \nu}
 {\Gamma (\half+\nu -\frac{i \omega }{2 \pi T }+i q e_d )
 \over \Gamma \left(\frac{1}{2}-\nu -\frac{i \omega }{2 \pi T}+i
q e_d \right)} \buildrel{T\to0}\over{\to} c_k \omega^{2\nu}  \ee
%
%\redcom{
%There is a numerical instability for $ \Im \omega < -\pi T$} \\
%\redcom{(can also be seen directly from the wave equation:} \\
%\redcom{the outgoing solution near the horizon dominates exponentially over the
%desired incoming solution.) }
We first note from this formula
\eqref{eq:sigT} that what
at $T=0$ was a branch cut for $\omega^{2\nu}$
has become a line of discrete poles of the Gamma function.

An interesting phenomenon which occurs at $T=0$ is that
as $k \to k_F$,
the quasiparticle pole sometimes moves under the branch cut and escapes onto another sheet of the complex
$\omega$ plane;
this leads to an extreme particle-hole asymmetry (visible in Fig 2C of \cite{Faulkner:2010da}).
%\footnote{
More precisely, when the pole hits the branch point of $\sG(\omega)$ at $\omega=0$,
it changes its `velocity' $d\omega/dk$ according to
$ v_F(|\vec k| - k_F) = \sG(\omega)$.
Depending on the phase of $\sG$,
this can put the pole on another sheet of the complex omega plane.
%}

This raises an interesting question: what happens to the pole which
at $T=0$ would have gone under the branch cut?
The answer is that it joins the line of poles approximating the branch cut;
their combined motion mimics the effects of the motion of the pole on the
second sheet.
This effect is visible in the sequence of pictures in Fig.~\ref{complex_omega}.

%[the behavior of the residue.]

At finite $T$, the pole no longer hits the real axis;
the distance of closest approach is of order $ (\pi T)^{2\nu} $.
This is thermal smearing of the Fermi surface.

\section{Discussion: charged AdS black holes and frustration}

The state we have been studying has a large low-lying density of states.
The entropy density of the black hole is
%$$ S_{BH} = {A \over 4 G_N} = N^{3/2} \CV \alpha^2 $$
\be s(T=0) = {1\over V_{d-1}} {A \over 4 G_N}= 2 \pi e_d \rho~.
\ee
At leading order in $1/N^2$, this is a large groundstate degeneracy.
The state we are studying is not supersymmetric, and so we
expect this degeneracy to be lifted at finite $N$.
The most likely way in which the third law of thermodynamics
will be enforced is by an instability to a superconducting state,
the implementation and effects of which are discussed below in \S\ref{sec:sc}.
In the absence of the necessary ingredients for a holographic superconductor,
one way in which the degeneracy can be lifted
was pointed out in \cite{Hartnoll:2009ns}.
%there is a finite density of fermions in the bulk.
The groundstate of the spinor field in the Reissner-Nordstr\"om (RN) black hole is a
Fermi sea of filled negative-energy states (a related discussion for a different black hole
appears in \cite{Hartman:2009qu}).
An arbitrarily small density of matter
close enough to the extremal horizon will produce
an order-one back-reaction on the bulk geometry.
According to \cite{Hartnoll:2009ns},
the density of spinor particles themselves
modify the far-IR of the $AdS_2\times \RR^2$ to a Lifshitz geometry
\cite{Kachru:2008yh}
with dynamical exponent $ z \propto N^2$.
This gravitating object, supported by the degeneracy pressure of charged fermions,
has been called an `electron star' \cite{electronstar}.
The modification occurs out to a very small distance from the horizon that scales like $e^{ - N^2}$,
and does not change the features of our calculations
that we have emphasized above.
(More concretely, the results are unchanged down to temperatures
scaling like $e^{-N^2}$.)
If one considers instead spinor fields whose mass and charge grows with $N$ (\eg\ like $q \sim N$),
then the back-reaction of a finite density can modify the geometry out to a larger radius
of order $1/N$
to a Lifshitz background with a dynamical exponent which is finite in the large $N$ limit \cite{electronstar}.

%Can we get this behavior without the large low-$E$ density of states?
%Presumably: The small-frequency behavior depended on {\it existence} of an IR CFT,
%and not on its large central charge
%$c \propto s(T=0)$ of IR CFT.

We should comment in more detail on the effects on $G_R$ of
changes to the near-horizon geometry resulting from the gravitational back-reaction of the bulk fermion
density\cite{Hartnoll:2009ns, Faulkner:2010tq}.
The geometry $AdS_2 \times \RR^2$ discussed above
is the $z\to \infty$ limit of the following family of metrics with Lifshitz scaling $t \to \lambda^z t, \vec x \to \lambda \vec x$:
 \be
 ds^2 =  {R^2_2 \ov \ze^2} \( d\tau^2 +
d \ze^2\)
 + {r_*^2 \ov R^2} \ze^{2/z} d \vec x^2
 \ee
Any finite $z$ modifies the non-analytic behavior of the self-energy \cite{Faulkner:2010tq}:
The Lifshitz scaling implies that the IR CFT scaling function takes the form $\sG_R(\omega, k) = \omega^{2\nu} F\( \frac{\omega}{k^z}\)$; note that it is $k$ and not $k_\perp$ that
appears in the scaling function -- the IR CFT knows nothing of $k_F$.
Lifshitz wave equations for general $z$ seem not to be exactly solvable, but
at small frequency a WKB analysis can be used to study the scaling function more explicitly \cite{Faulkner:2010tq};
the self-energy goes like $\exp{\(-1/\omega^{{1\over z-1}} \)}$ for frequencies less
than the scale at which the RN groundstate degeneracy is split.
We emphasize that nevertheless the non-Fermi liquid behavior persists down to
parametrically low energies (of order $\mu e^{-z}$).

%Many other resolutions of the `RN entropy puzzle' are possible.
\HL{There exist other geometries in which the ground entropy vanishes.}
For example,
the inclusion of other light bulk modes (\eg\ neutral scalars)
\cite{Herzog:2009gd, Gubser:2009qt, Goldstein:2009cv}
can have an important effect on the groundstate.
A systematic exploration of the fermion response
in the general holographic description of finite density
will be valuable.

\subsection{Fermions in the bulk and the oscillatory region}

In this brief subsection we observe that some features of the `electron star'
are already visible in the fermion correlators computed in the RN black hole at leading order in $1/N$.

The electronic states in the curved background correspond to the solutions of the bulk Dirac equation with normalizable boundary conditions in the UV and regular boundary conditions (ingoing in Lorentzian signature) in the IR. The transverse momenta can be chosen continuously, $k_x, k_y \in \IR$. In the radial $r$ direction, there is no translational invariance, so in this direction we describe the wavefunction in real space. The radial quantum number $n$ is discrete because boundary conditions have been specified.
Finally, after fixing the transverse momenta, the frequency should be adjusted such that the solution satisfies the boundary conditions. This happens for $\omega$ values where the retarded Green's function has a pole.

Instead of labeling the states by $\{ k_x, k_y, n \}$ where $n$ is the discrete radial quantum number, we can label them by $\{ k_x, k_y, \omega\}$ where $\omega$ is chosen from the discrete set of poles of the retarded correlator with fixed $k_x, k_y$. The states will be automatically filled for $Re \, \omega < 0$ and empty for $Re \, \omega > 0$.

If there were translational invariance in the radial direction, then $n$ would be a continuous parameter. When labeling the states by $\omega$, this means that there is a continuum of poles in the correlator from which we can choose from. (In practice, the poles would probably form a branch cut.) In the bulk, \DV{deep inside the electron star there is locally} a 3d Fermi surface. Since AdS/CFT should describe the excitations of the star, \DV{we expect infinitely many poles} in the Green's function \DV{for some values of the momenta}.

%(this is well-defined in a WKB limit)

Surprisingly, something like this already happens in the oscillatory region ($k<k_o$) in the Reissner-Nordstr\" om case.
This is a regime of momenta where the Green's functions are observed to have log-periodic
behavior in frequency, and where the zero-frequency spectral weight is nonvanishing \cite{Liu:2009dm}.
The basic observation is that although the poles do not yet form a branch cut, they do accumulate near $\omega=0$.
In tortoise coordinates one can see that exactly in this oscillatory region, the wavefunction ``falls into the black hole'', \ie\ it is localized in the IR.

%\DV{removed sentences}

%If we forget about backreaction, then these states are naturally identified with the states that make up the electron %star. This also means that without backreaction, the (small) electron star has its 3d Fermi surface at $k=k_\textrm{star} %= k_o$.

We note further that in examples without an oscillatory region (as is the case for the alternative quantization, see
the right part of Figure 5 of \cite{Faulkner:2009wj}),
there is no bulk Fermi sea near the horizon, and therefore no such modification of the geometry.
We must observe, however, that in known examples that have a Fermi surface but no oscillatory
region, there is a relevant operator, perturbation by which removes the Fermi surface.  It would be
interesting to find a stable example without an oscillatory region.

%
% We can interpret the series of poles as fractionalization of electrons into infinitely many pieces, corresponding to the radial modes.

%
%{pessimism:}
%\greencom{$S(T=0) \neq 0$ violates third law of thermodynamics, unphysical, weird string-theorist nonsense.}

%

%
%{optimism:}
%\greencom{this is a model of frustration:
%%\includegraphics[scale=0.2]{pictures/triangular_frustration.eps}  \ttref{[Wikipedia]}\\
%large degeneracy in classical limit \\
%%states whose energies differ by $N^\delta$ with $ \delta < 2$ can contribute.\\
%quantumly, very many states whose energy differences are parametrically smaller
%than some UV scale ($\mu$).}

%1. \textcolor{darkblue}{Third law:}

\section{Transport}

The most prominent mystery of the strange metal phase
is the linear-in-$T$ electrical resistivity.
%\begin{center}
%%\includegraphics[height=69pt]{pictures/tasi6}
%\includegraphics[height=69pt]{pictures/linear_T}
%%~~~\ttref{fig from [Polchinski, hep-th/9210046]}
%\end{center}
Electron-electron scattering (combined with umklapp or impurities) produces $\rho \sim T^2$,
electron-phonon scattering produces $\rho \sim T^5$.
A van Hove singularity requires fine tuning of the Fermi level.
No simple, robust effective field theory gives $\rho \sim T$.

AdS/CFT techniques are well-adapted for studying transport.
Holographic conductivity is computed by solving Maxwell's equations in the bulk.  The answer is
of the form
\be%{\rm Kubo:}~~~
\displaystyle{\sigma^{DC}   =\lim_{\omega\to 0}
\Im {1\over \omega} \vev{j^x j^x } (\omega, \vec 0)  } =O( \textcolor{darkred}{N^2})% (\text{boring})
+
 \textcolor{darkred}{N^0} \sigma_{{\rm from~spinor}}+ ...\ee
The \HL{$O(N^2)$ contribution does not know about $k_F$ as it only depends on the black hole geometry, and not on the dynamics of the fermions.}
The contribution to the conductivity from the holographic Fermi surface is down by $N^{-2}$:
its extraction requires a (spinor) loop in the bulk, as in \cite{Denef:2009yy, Denef:2009kn, CaronHuot:2009iq}.
\begin{figure}[h]
\begin{center}
\includegraphics[height=60pt]{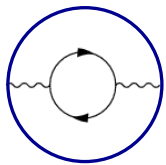}
\caption{The `Witten diagram' which gives the holographic Fermi surface contribution
to the resistivity.  The wiggly lines represent bulk gauge fields
propagating from the boundary of AdS (dark blue circle);
the solid lines represent bulk spinor propagators.}
\label{fig:pol}
\end{center}
\end{figure}

%\subsection{Cartoon of conductivity calculation}

\def\vertexZ{\Lambda}

The result is very similar to the calculation of Fermi liquid conductivity
(\eg\ \cite{Mahan}), with extra integrals over $r$,
and no vertex corrections.
The only process that contributes to leading order
is depicted in Fig.~\ref{fig:pol}.
The key step is to relate the bulk spinor spectral function
to that of the boundary fermion operator:
\be \Im S_{\alpha\beta}(\Omega, k; r_1, r_2) =
{\psi^b_\alpha(\Omega, k, r_1) \bar \psi^b_\beta(\Omega, k, r_2) \over W_{ab} } A(\Omega,k)
\ee
where
$A = {1\over \pi} \Im G_R$ is the spectral weight computed above.
The conductivity is
\be \sigma^{DC} =  S_d \int dk k^{d-2}
\int d\omega {df \over d\omega} \vertexZ^2(k, \omega) A(\omega, k)^2 \ee
$f$ is the fermi function, and
$\vertexZ \sim q \int_{r_0}^\infty dr \sqrt g g^{xx} A_x(r,0) {\bar\psi^b (r) \Gamma^x \psi^b(r)\over W_{ab} } $
is a vertex factor, encoding
data analogous to the UV coefficients $v_F, h_{1,2}$ in \eqref{eq:GR}.

\be \sigma^{DC}_{\rm from~FS} = \Upsilon^2 k_F^{d-2} \int d\omega {f'(\omega) \over
% \tilde h_2
 \Im \sG(\omega)}
 \sim T^{- 2 \nu} \ee
The factor $\Upsilon$ is determined by boundstate wavefunctions,
$k_F^{d-2}$ is the volume of the Fermi surface.
In the last step we have used the scaling relation
$ \Im \sG\( T\to 0, \omega/T~{\rm fixed}\)= T^{2\nu} g(\omega/T) $.

In the marginal Fermi liquid case where $\nu=\half$,
this indeed gives
\be \rho = \(\sigma^{DC}\)^{-1} \sim T~. \ee
In contrast to the models of NFL that arise by coupling
a Fermi surface to a gapless bosonic mode,
the transport lifetime and the single-particle lifetime have the
same temperature dependence.
This is possible because the quasiparticle decay
is mediated by the IR CFT,
which contains low-energy modes for nonzero (indeed, for any) momentum.

Although the DC resistivity is not sensitive to the stability
of the quasiparticles, the behavior
of the optical conductivity does change dramatically at $\nu = \half$~\cite{Faulkner:2010da, resistivitypaper}.
%we refer the reader to \cite{Faulkner:2010da, resistivitypaper} for a discussion of this.

\section{The superconducting state}
\label{sec:sc}

A useful test of any model of the normal state
is whether it can incorporate the transition to
superconductivity.
It is indeed possible to describe superconductivity
holographically by
including charged scalar fields in the black hole background.
At low temperature, they can condense,
spontaneously breaking the $U(1)$ symmetry,
changing the background \cite{holographicsc1,holographicsc2}.

%
%\begin{wrapfigure}{r}{45mm}
%\vspace{-10pt}
%\includegraphics[height=40pt]{pictures/high_tc.eps}~~
%\includegraphics[scale=0.3]{pictures/ARPES_antinode}
%%\caption{\label{fig:buoyancy}}
%\vspace{-20pt}
%\end{wrapfigure}
%%\bl{2.}

% \ttref{[Hartnoll et al, Gubser 2008]}.

The problem of the fermion response in various possible
holographic superconducting phases has been studied in
\cite{chenkaowen, Faulkner:2009am, fabio, Gubser:2010dm, Ammon:2010pg, Benini:2010qc, Vegh:2010fc}.
The superconducting condensate can open a gap in the fermion spectrum
around the chemical potential
\cite{Faulkner:2009am} if a suitable bulk coupling between
spinor and scalar is included.
\def\CC{C}
The bulk action we consider for the fermion is
\be
\label{majoranaaction}
S[\spinor] = \int d^{d+1}x
\sqrt{-g}
\[  i \bar \spinor \( \Gamma^M D_M  - \mspinor  \)\spinor
+ \eta_5^\star \scalar^\star  \spinor^T  \CC  \Gamma^5 \spinor + \eta_5 \scalar \bar \spinor \CC \Gamma^5  \bar \spinor^T
\]~~.
\ee
$\scalar$ is the scalar field whose condensation spontaneous breaks
the $U(1)$ symmetry.
$\CC$ is the charge conjugation matrix,
and $\Gamma^5$ is the chirality matrix, $ \{ \Gamma^5, \Gamma^M\} = 0$.
The superconducting order parameter in these studies is s-wave, but is not BCS.

\begin{figure}[h!]
\begin{center}
\includegraphics[scale=.75]{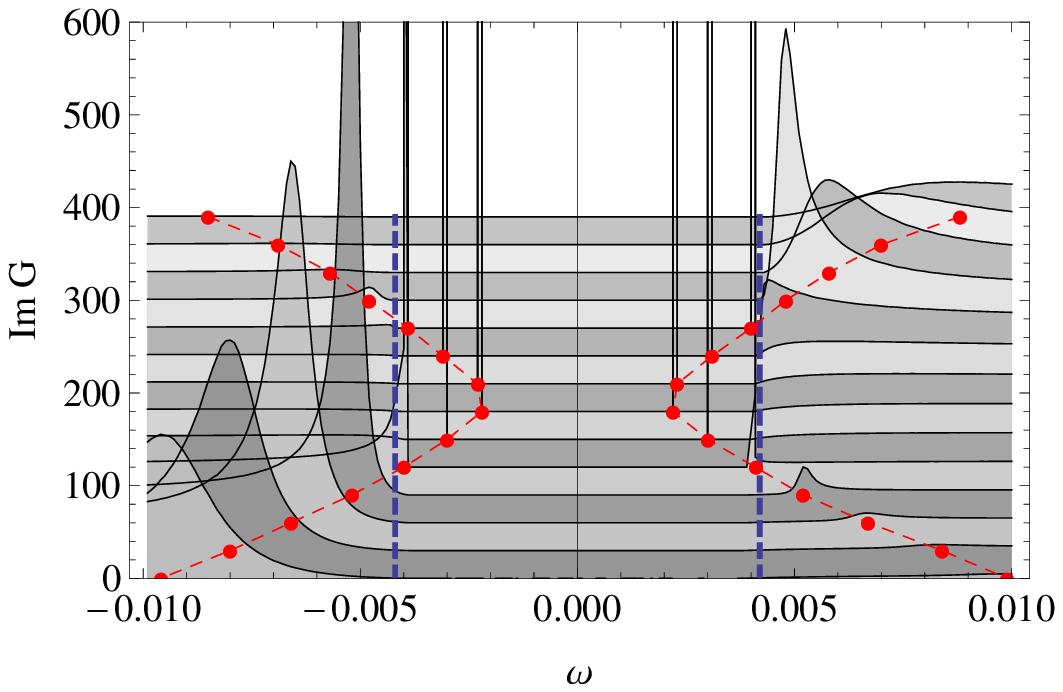}
\caption{The effect of the
%Majorana coupling
superconducting order
on the fermion spectral density (from \cite{Faulkner:2009am}). Shown are plots
of $A(k, \omega)$ at various $k \in [.81, .93]$ for $\qspinor = \half, \mspinor = 0 $ in a
low-temperature background of a scalar with $ \qscalar = 1, \mscalar^2 =-1$ (first constructed in \cite{Horowitz:2009ij}), with $\eta_5 = 0.025$. The blue dashed line indicates the boundary of the region in
which the incoherent part of the spectral density is completely suppressed, and the lifetime of
the quasiparticle is infinite. The red dotted line indicates the location of the peak.
 \label{fig:gap3}}
\end{center}
\end{figure}

Interestingly, in the condensed phase, one finds stable quasiparticles, even when $\nu \leq \half$.  The modes into which the quasiparticle decays in the normal state
are lifted by the superconducting condensate.
The quasiparticles are stable in a certain kinematical regime,
similar to Landau's critical velocity for drag in a superfluid:
these holographic superconductor groundstates
do have gapless excitations (in fact a relativistic CFT worth),
but the Fermi surface can occur outside their lightcone
(visible as the blue dashed line in Fig.~\ref{fig:gap3}).

%\end{frame}\begin{frame}{Other groundstates}

%\vskip-.2in
%\onslide<1,2,3>
%\textcolor{darkblue}{Charged bosons:} In many explicit dual pairs, $\exists$ charged scalars.
%

%$\bullet$ If their mass/charge is big enough, they don't condense \cite{Denef:2009tp}:
%\redcom{(vs: a weakly-coupled charged boson at $\mu \neq 0$ will condense.)}
%\begin{wrapfigure}{r}{40mm}
%\vspace{-40pt}
%\includegraphics[scale=0.25]{pictures/stability_diagram}
%%\caption{\label{fig:buoyancy}}
%\vspace{-40pt}
%\end{wrapfigure}
%\greencom{Like moduli stabilization.}
%\hskip1in
%\begin{center}\includegraphics[height=100pt]{pictures/stability_diagram}
%~~~\ttref{[Denef-Hartnoll]}
%\end{center}
%\vskip-.1in

%
%%\onslide<2,3>
%\begin{wrapfigure}{r}{40mm}
%\vspace{-20pt}
%\includegraphics[scale=0.2]{pictures/hussey.eps}
%%\caption{\label{fig:buoyancy}}
%\vspace{-25pt}
%\end{wrapfigure}
%\vskip.2in
%%\onslide<1,2,3>
%$\bullet$ Many systems to which we'd like to apply this also have a superconducting region.
%%\vskip-2in
%%\begin{flushright}
%%\includegraphics[scale=0.10]{pictures/hussey.eps}
%%\end{flushright}

%\onslide<3>

%Other light bulk modes (\eg\ neutral scalars)
%can also have
%an important effect on the groundstate
%\cite{Herzog:2009gd, Gubser:2009qt, Goldstein:2009cv}.

\section{Conclusions}

%
%{Other interesting features}
%\begin{itemize}
%\item particle-hole asymmetry
%\item disappearance of FS under relevant deformation of UV CFT
%\item if $\nu \in i \IR ~\longrightarrow~~$ $G$ is log-periodic in frequency
%\item $\exists$ formula for Fermi velocity in terms of boundstate wavefunctions
%\item $\exists$ no free-fermion limit
%\item statistics and stability
%\item branch point resolved at finite temperature
%\end{itemize}

\HL{The holographic calculation of
the spectral function can be described by a simple low energy effective theory}
\cite{Faulkner:2009wj, Faulkner:2010tq, Faulkner:2010da, Faulkner:2010jy}.
Consider a Fermi liquid (with creation operator $\psi$)
{\greent{mixing}} with a {\redt{bath}} of critical {\greent{fermionic}} fluctuations
{\bl{with large
dynamical exponent}}:
%\ttref{[FLMV 0907.2694, Faulkner-Polchinski 1001.5049, FLMV+Iqbal 1003.1728]}
\vskip-.1in
\be L = \bar \psi \( \omega - v_F k \) \psi
+ \bar \psi {\redt{\chi}}  + \psi \bar {\redt{\chi}} +
{\redt{\bar \chi \sG^{-1} \chi}} \ee
\redt{$\chi$} is an operator in the IR CFT;
the $k$-independence of its correlations suggests
that it arises from {\it localized} critical degrees of freedom.
The corrected $\psi$ Green's function is given by the geometric series:
\vskip-.3in
\begin{fmffile}{fmfp46}
\begin{equation*}
\parbox{17mm}{\begin{fmfgraph*}(50,30)
\fmfleft{i1}
\fmfright{o1}
\fmf{plain, tension=.7}{i1,v1}
\fmfblob{.5w}{v1}
\fmf{plain, tension=.7}{v1,o1}
\end{fmfgraph*}}
=~
\parbox{13mm}{\begin{fmfgraph*}(30,15)
\fmfleft{i1}
\fmfright{o1}
\fmf{plain}{i1,o1}
\end{fmfgraph*}}
+
\parbox{22mm}{
\begin{large}
\begin{fmfgraph*}(60,60)
\fmfleft{i1}
\fmfright{o1}
\fmf{plain, tension=.7}{i1,v1}
\fmf{dbl_plain, tension=.6}{v1,v2}
\fmf{plain, tension=.7}{v2,o1}
\fmfdot{v1,v2}
\end{fmfgraph*}
\end{large}}
+
\parbox{38mm}{
\begin{large}\begin{fmfgraph*}(100,70)
\fmfleft{i1}
\fmfright{o1}
\fmf{plain, tension=.7}{i1,v1}
\fmf{dbl_plain, tension=.5}{v1,v2}
\fmf{plain, tension=.5}{v2,v3}
\fmf{dbl_plain, tension=.5}{v3,v4}
\fmf{plain, tension=.7}{v4,o1}
\fmfdot{v1,v2,v3,v4}
\end{fmfgraph*}
\end{large}}
+ \dots
\end{equation*}
\end{fmffile}
\vskip-.4in
\be \vev{ \bar \psi \psi}
= { 1 \over \omega - v_F k - \redt{\sG} }~~~~~~~
\redt{ \sG = \vev{\bar \chi \chi} = c(k) \omega^{2\nu} } .\ee
Note that according to the free fermion scaling, $\redt{\chi}$ has dimension zero;
therefore for $\nu \leq \half$, the $\bar \psi \redt{\chi} $ coupling is a relevant perturbation.

Above, we have described a certain class of fixed points
which have some features in common with various non Fermi liquid
metals.
We find Fermi surfaces with vanishing quasiparticle residue.
The single-fermion self-energy is a power-law in frequency,
independent of momentum, as in DMFT and some slave-particle
descriptions.
Whether the states we describe can arise from any model
of electrons with short-range interactions is an important open question\footnote{
\HL{See~\cite{Sachdev:2010um,qimiao} for some recent ideas along this direction.}}.
%has argued that the answer is `yes'.}.

%
%\begin{enumerate}
%\item{} The green's function near the FS is of the form
%found previously in perturbative calculations, but the
%nonanalyticity can be order one.

%%\item{} The RG matching between the UV CFT and the IR CFT
%%can be modelled as a coupling:
%%$$
%% S = \int d\om d\vk \,
%%  \sO_{UV}^\da \, \Sig_{\om, \vk_\perp}\,  \sO_{UV} +
%%\int_{FS}
%%  \bar D_{\om, \vk_F} \sO_{UV}^\da \sO_{IR} + h.c.
%%$$
%% $\Sig, D$ are real, analytic in $\omega$, UV data
%% %the correlation of $\sO_I (\om, \vk_F)$ is controlled by an IR CFT
%%$
%%\vev{\sO_{IR} (\om, \vk_F)^\da \sO_{IR} (\om, \vk_F')}_{IR}
%%= \sG_{k_F} (\om) \delta_{\vk_F, \vk_F'} \ .
%%$
%%$$
%%\Longrightarrow ~
%%G_R (\om, \vk_\perp)
%%\equiv \vev{\sO_{UV}^\da \sO_{UV}}
%% = {1 \ov \Sig (\om, \vk_\perp) + D^2 (\om, \vk_F) \sG_{k_F} (\om)}
%% $$

%\item{} The knowledge of quantum statistics displayed by
%the classical wave equations is remarkable and
%necessary for AdS/CFT to be consistent with basic facts about
%many-body physics.

%
%\item{}  The leading $N^{-1}$
%contribution to the free energy
%exhibits quantum oscillations in a magnetic field \cite{Denef:2009yy}.

%\item{} We should compute the specific heat.  \cite{tomnabil}

%\end{enumerate}

%

%\end{document}

\begin{acknowledgements}
We thank T.~Senthil for many useful discussions,
and G.~Horowitz and M.~Roberts for collaboration on \cite{Faulkner:2009am}.
Work supported in part by funds provided by the U.S. Department of Energy
(D.O.E.) under cooperative research agreement DE-FG0205ER41360 and the OJI program,
and in part by the National Science Foundation under Grant No. NSF PHY05-51164.
The work of JM is supported in part by an Alfred P. Sloan Fellowship.

\end{acknowledgements}


\begin{thebibliography}{60}

\setlength{\itemsep}{1.2pt}
\providecommand{\natexlab}[1]{#1}
\expandafter\ifx\csname urlstyle\endcsname\relax
  \providecommand{\doi}[1]{doi:\discretionary{}{}{}#1}\else
  \providecommand{\doi}{doi:\discretionary{}{}{}\begingroup
  \urlstyle{rm}\Url}\fi




\bibitem[{Affleck(1988)}]{affleck}
I. Affeck, in {\it Fields, Strings and Critical Phenomena}, p. 563-640, (ed. E. Brzin and J. Zinn-Justin North-Holland, Amsterdam, 
1990). 


\bibitem[{Aharony \emph{et~al.}(2000)}]{magoo}
  O.~Aharony, S.~S.~Gubser, J.~M.~Maldacena, H.~Ooguri and Y.~Oz,
  ``Large N field theories, string theory and gravity,''
  Phys.\ Rept.\  {\bf 323}, 183 (2000)
  [arXiv:hep-th/9905111].


\bibitem[{Altshuler \emph{et~al.}(1994)}]{altshuler:1994}
B.~L.~Altshuler, L.~B.~Ioffe and A.~J.~Millis,
``On the low energy properties of fermions with singular interactions,"
arXiv:cond-mat/9406024.

\bibitem[{Ammon \emph{et~al.}(2010)}]{Ammon:2010pg}
  M.~Ammon, J.~Erdmenger, M.~Kaminski and A.~O'Bannon,
  ``Fermionic Operator Mixing in Holographic p-wave Superfluids,''
  JHEP {\bf 1005}, 053 (2010)
  [arXiv:1003.1134 [hep-th]].


\bibitem[{Baym  \emph{et~al.}(1990)}]{baym}
G. Baym, H. Monien, C. J. Pethick, and D. G. Ravenhall,
``Transverse interactions and transport in relativistic quark-gluon and electromagnetic plasmas,"
Phys. Rev. Lett. {\bf 64} (1990) 1867.


%\cite{Benfatto:1990zz}
\bibitem[{Benfatto \& Gallavotti(1990)}]{Benfatto:1990zz}
G.~Benfatto and G.~Gallavotti,
``Renormalization-Group Approach to the Theory of the Fermi Surface,''
Phys.\ Rev.\  B {\bf 42} (1990) 9967.
%%CITATION = PHRVA,B42,9967;%%


\bibitem[{Benini \emph{et~al.}(2010)}]{Benini:2010qc}
  F.~Benini, C.~P.~Herzog and A.~Yarom,
  ``Holographic Fermi arcs and a d-wave gap,''
  arXiv:1006.0731 [hep-th].
  

\bibitem[{Birmingham \emph{et~al.}(2002)}]{Birmingham:2001pj}
  D.~Birmingham, I.~Sachs and S.~N.~Solodukhin, ``Conformal field theory interpretation of black hole quasi-normal modes,''
  Phys.\ Rev.\ Lett.\  {\bf 88}, 151301 (2002)
  [arXiv:hep-th/0112055].


\bibitem[{Boyanovsky \& de Vega(2001)}]{Boyanovsky}
  D.~Boyanovsky and H.~J.~de Vega,
  ``Non-Fermi liquid aspects of cold and dense QED and QCD: Equilibrium and
  non-equilibrium,''
  Phys.\ Rev.\  D {\bf 63}, 034016 (2001)
  arXiv:hep-ph/0009172;
  %%CITATION = PHRVA,D63,034016;%%


\bibitem[{Caron-Huot \& Saremi(2009)}]{CaronHuot:2009iq}
S.~Caron-Huot and O.~Saremi,
``Hydrodynamic Long-Time Tails from Anti De Sitter Space,''
arXiv:0909.4525 [hep-th].

\bibitem[{Chen \emph{et~al.}(2009)}]{chenkaowen}
J.~W.~Chen, Y.~J.~Kao and W.~Y.~Wen,
``Peak-Dip-Hump from Holographic Superconductivity,''
arXiv:0911.2821 [hep-th].



%\cite{Cubrovic:2009ye}
\bibitem[{Cubrovic \emph{et~al.}(2009)}]{Cubrovic:2009ye}
  M.~Cubrovic, J.~Zaanen and K.~Schalm,
  ``Fermions and the AdS/CFT correspondence: quantum phase transitions and the
  emergent Fermi-liquid,''
  arXiv:0904.1993 [hep-th].
  
  

\bibitem[{Damascelli \emph{et~al.}(2003)}]{damascelli}
A.~Damascelli, Z.~Hussain, Z-X.~Shen,
``Angle-resolved photoemission studies of the cuprate superconductors,''
Rev.\ Mod.\ Phys.\ {\bf 75}, 473 - 541 (2003).



\bibitem[{Denef \emph{et~al.}(2007)}]{modulistabilization2}
    F.~Denef, M.~R.~Douglas and S.~Kachru,
  ``Physics of string flux compactifications,''
  Ann.\ Rev.\ Nucl.\ Part.\ Sci.\  {\bf 57}, 119 (2007)
  [arXiv:hep-th/0701050];



\bibitem[{Denef(2008)}]{modulistabilization5}
F.~Denef,
``Les Houches Lectures on Constructing String Vacua,'' arXiv:0803.1194 [hep-th].


\bibitem[{Denef \emph{et~al.}(2009a)}]{Denef:2009yy}
F.~Denef, S.~A.~Hartnoll and S.~Sachdev,
``Quantum Oscillations and Black Hole Ringing,''
arXiv:0908.1788 [hep-th].

\bibitem[{Denef \emph{et~al.}(2009b)}]{Denef:2009kn}
    F.~Denef, S.~A.~Hartnoll and S.~Sachdev,
  ``Black Hole Determinants and Quasinormal Modes,''
arXiv:0908.2657 [hep-th].


\bibitem[{Douglas \& Kachru(2006)}]{modulistabilization3}
    M.~R.~Douglas and S.~Kachru,
  ``Flux compactification,''
  Rev.\ Mod.\ Phys.\  {\bf 79}, 733 (2007)
  [arXiv:hep-th/0610102];


%\cite{Faulkner:2009wj}
\bibitem[{Faulkner \emph{et~al.}(2009a)}]{Faulkner:2009wj}
T.~Faulkner, H.~Liu, J.~McGreevy and D.~Vegh,
``Emergent Quantum Criticality, Fermi Surfaces, and Ad$S_2$,''
arXiv:0907.2694 [hep-th];



%\cite{Faulkner:2009am}
\bibitem[{Faulkner \emph{et~al.}(2009b)}]{Faulkner:2009am}
T.~Faulkner, G.~T.~Horowitz, J.~McGreevy, M.~M.~Roberts and D.~Vegh,
``Photoemission `Experiments' on Holographic Superconductors,''
arXiv:0911.3402 [hep-th].
%%CITATION = ARXIV:0911.3402;%%



\bibitem[{Faulkner \& Polchinski(2010)}]{Faulkner:2010tq}
  T.~Faulkner and J.~Polchinski,
  ``Semi-Holographic Fermi Liquids,''
  arXiv:1001.5049 [hep-th].



\bibitem[{Faulkner \emph{et~al.}(2010a)}]{Faulkner:2010da}
  T.~Faulkner, N.~Iqbal, H.~Liu, J.~McGreevy and D.~Vegh,
  ``Strange Metal Transport Realized by Gauge/Gravity Duality"
%  ``From black holes to strange metals,''
  {\bf Science}, % 27 August 2010:
Vol. 329. no. 5995 (2010) 1043
[arXiv:1003.1728 [hep-th]].




\bibitem[{Faulkner \emph{et~al.}(2010b)}]{resistivitypaper}
T.~Faulkner, N.~Iqbal, H.~Liu, J.~McGreevy and D.~Vegh,
``Charge transport by holographic non-Fermi liquids"
to appear.


  \bibitem[{Faulkner \emph{et~al.}(2010c)}]{Faulkner:2010jy}
  T.~Faulkner, H.~Liu and M.~Rangamani,
  ``Integrating out geometry: Holographic Wilsonian RG and the membrane
  paradigm,''
  arXiv:1010.4036 [hep-th].




\bibitem[{Festuccia \& Liu(2006)}]{Festuccia:2005pi}
  G.~Festuccia and H.~Liu,
  ``Excursions beyond the horizon: Black hole singularities in Yang-Mills
  theories. I,''
  JHEP {\bf 0604}, 044 (2006)
  [arXiv:hep-th/0506202].




\bibitem[{Freedman \emph{et~al.}(1999)}]{Freedman:1998tz}
  D.~Z.~Freedman, S.~D.~Mathur, A.~Matusis and L.~Rastelli,
  ``Correlation functions in the CFT($d$)/AdS($d+1$) correspondence,''
  Nucl.\ Phys.\  B {\bf 546}, 96 (1999)
  [arXiv:hep-th/9804058].


\bibitem[{Goldstein \emph{et~al.}(2009)}]{Goldstein:2009cv}
K.~Goldstein, S.~Kachru, S.~Prakash and S.~P.~Trivedi,
``Holography of Charged Dilaton Black Holes,''
arXiv:0911.3586 [hep-th].


\bibitem[{Grana(2005)}]{modulistabilization4}
M.~Grana,
 ``Flux compactifications in string theory: A comprehensive review,'' Phys.\ Rept.\ {\bf 423}, 91 (2006) [arXiv:hep-th/0509003];   %%CITATION = PRPLC,423,91;%%




\bibitem[{Gubser \emph{et~al.}(1998)}]{AdS/CFT3}
S.~S.~Gubser, I.~R.~Klebanov and A.~M.~Polyakov,
``Gauge theory correlators from non-critical string theory,''
Phys.\ Lett.\ B {\bf 428}, 105 (1998).
 

\bibitem[{Gubser(2008)}]{holographicsc1}
  S.~S.~Gubser,
 ``Breaking an Abelian gauge symmetry near a black hole horizon,''
  Phys.\ Rev.\  D {\bf 78}, 065034 (2008)
  [arXiv:0801.2977 [hep-th]];


\bibitem[{Gubser \& Rocha(2009)}]{Gubser:2009qt}
S.~S.~Gubser and F.~D.~Rocha,
``Peculiar Properties of a Charged Dilatonic Black Hole in $\mathrm{AdS}_5$,''
arXiv:0911.2898 [hep-th].

\bibitem[{Gubser \emph{et~al.}(2009)}]{fabio}
S.~S.~Gubser, F.~D.~Rocha and P.~Talavera,
``Normalizable Fermion Modes in a Holographic Superconductor,''
arXiv:0911.3632 [hep-th].


\bibitem[{Gubser \emph{et~al.}(2010)}]{Gubser:2010dm}
  S.~S.~Gubser, F.~D.~Rocha and A.~Yarom,
  ``Fermion correlators in non-abelian holographic superconductors,''
  arXiv:1002.4416 [hep-th].



 %\cite{Halperin:1992mh}
\bibitem[{Halperin \emph{et~al.}(1992)}]{Halperin:1992mh}
  B.~I.~Halperin, P.~A.~Lee and N.~Read,
  ``Theory of the half filled Landau level,''
  Phys.\ Rev.\  B {\bf 47}, 7312 (1993).
  %%CITATION = PHRVA,B47,7312;%%


\bibitem[{Hartman \& Strominger(2008)}]{Hartman:2008dq}
  T.~Hartman and A.~Strominger,
  ``Central Charge for $AdS_2$ Quantum Gravity,''
  JHEP {\bf 0904}, 026 (2009)
  [arXiv:0803.3621 [hep-th]].



\bibitem[{Hartman \emph{et~al.}(2009)}]{Hartman:2009qu}
  T.~Hartman, W.~Song and A.~Strominger,
  ``The Kerr-Fermi Sea,''
  arXiv:0912.4265 [hep-th].




\bibitem[{Hartnoll \emph{et~al.}(2008)}]{holographicsc2}
  %%CITATION = PHRVA,D78,065034;%%
  S.~A.~Hartnoll, C.~P.~Herzog and G.~T.~Horowitz,
%  ``Building a Holographic Superconductor,''
  Phys.\ Rev.\ Lett.\  {\bf 101}, 031601 (2008)
  [arXiv:0803.3295 [hep-th]],
%  S.~A.~Hartnoll, C.~P.~Herzog and G.~T.~Horowitz,
 % ``Holographic Superconductors,''
  JHEP {\bf 0812}, 015 (2008)
  [arXiv:0810.1563 [hep-th]];
  
  \bibitem[{Hartnoll(2009a)}]{Hartnoll:2009sz}
S.~A.~Hartnoll,
``Lectures on Holographic Methods for Condensed Matter Physics,''
arXiv:0903.3246 [hep-th].


\bibitem[{Hartnoll(2009b)}]{Hartnoll:2009qx}
S.~A.~Hartnoll,
``Quantum Critical Dynamics from Black Holes,''
arXiv:0909.3553 [cond-mat.str-el].


\bibitem[{Hartnoll \emph{et~al.}(2009)}]{Hartnoll:2009ns}
S.~A.~Hartnoll, J.~Polchinski, E.~Silverstein and D.~Tong,
``Towards Strange Metallic Holography,''
arXiv:0912.1061 [hep-th].


\bibitem[{Hartnoll \& Tavanfar(2010)}]{electronstar}
  S.~A.~Hartnoll and A.~Tavanfar,
  ``Electron stars for holographic metallic criticality,''
  arXiv:1008.2828 [hep-th].



\bibitem[{Herzog(2009)}]{holographicsc3}
  C.~P.~Herzog,
%  ``Lectures on Holographic Superfluidity and Superconductivity,''
  J.\ Phys.\ A  {\bf 42}, 343001 (2009)
  [arXiv:0904.1975 [hep-th]];
  %\cite{Horowitz:2010gk}


\bibitem[{Herzog \emph{et~al.}(2009)}]{Herzog:2009gd}
C.~P.~Herzog, I.~R.~Klebanov, S.~S.~Pufu and T.~Tesileanu,
``Emergent Quantum Near-Criticality from Baryonic Black Branes,''
arXiv:0911.0400 [hep-th].




\bibitem[{Holstein  \emph{et~al.}(1973)}]{Holstein:1973zz}
  T.~Holstein, R.~E.~Norton and P.~Pincus,
  ``de Haas-van Alphen Effect and the Specific Heat of an Electron Gas,''
  Phys.\ Rev.\  B {\bf 8}, 2649 (1973).
  %%CITATION = PHRVA,B8,2649;%%


\bibitem[{Horowitz(2010)}]{holographicsc4}
  G.~T.~Horowitz,
  %``Introduction to Holographic Superconductors,''
  arXiv:1002.1722 [hep-th].
  %%CITATION = ARXIV:1002.1722;%%
  %%CITATION = JPAGB,A42,343001;%%
  
  \bibitem[{Horowitz \& Roberts(2009)}]{Horowitz:2009ij}
G.~T.~Horowitz and M.~M.~Roberts,
``Zero Temperature Limit of Holographic Superconductors,''
JHEP {\bf 0911} (2009) 015
[arXiv:0908.3677 [hep-th]].


\bibitem[{Iqbal \& Liu(2009)}]{Iqbal:2009fd}
N.~Iqbal and H.~Liu,
``Real-Time Response in AdS/CFT with Application to Spinors,''
Fortsch.\ Phys.\  {\bf 57} (2009) 367
[arXiv:0903.2596 [hep-th]].




\bibitem[{Iqbal \emph{et~al.}(2010)}]{Iqbal:2010eh}
  N.~Iqbal, H.~Liu, M.~Mezei and Q.~Si,
  ``Quantum phase transitions in holographic models of magnetism and
  superconductors,''
  Phys.\ Rev.\  D {\bf 82}, 045002 (2010)
  [arXiv:1003.0010 [hep-th]].
  
  \bibitem[{Kachru \emph{et~al.}(2008)}]{Kachru:2008yh}
S.~Kachru, X.~Liu and M.~Mulligan,
``Gravity Duals of Lifshitz-Like Fixed Points,''
Phys.\ Rev.\  D {\bf 78} (2008) 106005
[arXiv:0808.1725 [hep-th]].





\bibitem[{Kim \emph{et~al.}(1994)}]{Patrick}
Y.~B.~Kim, A.~Furusaki, P.~A.~Lee, and X-G.~Wen,
``Gauge-invariant response functions of fermions coupled to a gauge field,"
Phys. Rev. B 50, 17917 (1994);
Y.~B.~Kim, P.~A.~Lee, and X-G.~Wen,
``Quantum Boltzmann equation of composite fermions interacting with a gauge field''
Phys. Rev. B 52, 17275 (1995).



\bibitem[{Lawler \emph{et~al.}(2006)}]{lawler}
M.~Lawler et al,
%``Non-perturbative behavior of the quantum phase transition to a nematic Fermi fluid"
Phys.\ Rev.\ {\bf B 73}, 085101 (2006)
[arXiv:cond-mat/0508747];
%```Local' Quantum Criticality at the Nematic Quantum Phase Transition"
M.~J.~Lawler, E.~Fradkin,
Phys.\ Rev.\ {\bf B 75}, 033304 (2007)
[arXiv:cond-mat/0605203].




\bibitem[{Lee \& Nagaosa(1992)}]{nagaosa}
P.~A.~Lee and N.~Nagaosa,
``Gauge theory of the normal state of high-Tc superconductors,''
Phys. Rev. B 46, 5621 (1992).


%\cite{Lee:2008xf}
\bibitem[{Lee(2008)}]{Lee:2008xf}
  S.~S.~Lee,
  ``A Non-Fermi Liquid from a Charged Black Hole: A Critical Fermi Ball,''
  arXiv:0809.3402 [hep-th].
  %%CITATION = ARXIV:0809.3402;%%


\bibitem[{Lee(2009)}]{sungsikgaugefield}
S.~S.~Lee,
``Low energy effective theory of Fermi surface coupled with U(1) gauge field in 2+1 dimensions,"
Phys.\ Rev.\ {\bf B 80}, 165102 (2009)
[arXiv:0905.4532 [cond-mat.str-el]].



\bibitem[{Liu \emph{et~al.}(2009)}]{Liu:2009dm}
H.~Liu, J.~McGreevy and D.~Vegh,
``Non-Fermi Liquids from Holography,''
arXiv:0903.2477 [hep-th].



\bibitem[{Mahan(2000)}]{Mahan}
G.~Mahan, {\it Many-Particle Physics}, Plenum Press.



\bibitem[{Maldacena(1998)}]{AdS/CFT1}
J.~M.~Maldacena,
``The large N limit of superconformal field theories and supergravity,''
Adv.\ Theor.\ Math.\ Phys.\  {\bf 2}, 231 (1998);


  \bibitem[{McGreevy(2009)}]{McGreevy:2009xe}
J.~McGreevy,
``Holographic Duality with a View Toward Many-Body Physics,''
 Adv.\ High Energy Phys.\  {\bf 2010}, 723105 (2010)
  [arXiv:0909.0518 [hep-th]].



%\cite{Metlitski:2010pd}
\bibitem[{Metlitski \& Sachdev(2010)}]{Metlitski:2010pd}
  M.~A.~Metlitski and S.~Sachdev,
  ``Quantum phase transitions of metals in two spatial dimensions: I.
  Ising-nematic order,''
  Phys.\ Rev.\  B {\bf 82}, 075127 (2010)
  [arXiv:1001.1153 [cond-mat.str-el]].
  %%CITATION = PHRVA,B82,075127;%%


%\cite{Mross:2010rd}
\bibitem[{Mross \emph{et~al.}(2010)}]{Mross:2010rd}
  D.~F.~Mross, J.~McGreevy, H.~Liu and T.~Senthil,
  ``A controlled expansion for certain non-Fermi liquid metals,''
  Phys.\ Rev.\ B {\bf 82}, 045121 (2010)
  [arXiv:1003.0894 [cond-mat.str-el]].
  %%CITATION = ARXIV:1003.0894;%%

\bibitem[{Nave \& Lee(2007)}]{nave}
C.~P.~Nave and P.~A.~Lee,
``Transport properties of a spinon Fermi surface coupled to a U(1) gauge field,"
Phys. Rev. B 76, 235124 (2007).


\bibitem[{Nayak \& Wilczek(1993)}]{Nayak:1993uh}
  C.~Nayak and F.~Wilczek,
  ``Non-Fermi liquid fixed point in (2+1)-dimensions,''
  Nucl.\ Phys.\  B {\bf 417}, 359 (1994)
  arXiv:cond-mat/9312086,
    ``Renormalization group approach to low temperature properties of a non-Fermi
  liquid metal,''
  Nucl.\ Phys.\  B {\bf 430}, 534 (1994)
  arXiv:cond-mat/9408016.




\bibitem[{Oganesyan \emph{et~al.}(2001)}]{fradkin}
V.~ Oganesyan, S.~Kivelson, E.~Fradkin,
``Quantum Theory of a Nematic Fermi Fluid,"
Phys.\ Rev.\ {\bf B 64}, 195109 (2001),
arXiv:cond-mat/0102093v2 [cond-mat.str-el].


%\cite{Polchinski:1992ed}
\bibitem[{Polchinski(1992)}]{Polchinski:1992ed}
J.~Polchinski,
``Effective Field Theory and the Fermi Surface,''
arXiv:hep-th/9210046.




\bibitem[{Polchinski(1993)}]{Polchinski:1993ii}
  J.~Polchinski,
  ``Low-energy dynamics of the spinon gauge system,''
  Nucl.\ Phys.\  B {\bf 422}, 617 (1994)   arXiv:cond-mat/9303037.
  %%CITATION = NUPHA,B422,617;%%2



\bibitem[{Reizer(1989)}]{reizer}
M. Y. Reizer,
``Relativistic effects in the electron density of states, specific heat, and the electron spectrum of normal metals,"
Phys. Rev. B {\bf 40}, 11571 (1989).

  
\bibitem[{Rey(2009)}]{soojong}
S.~J.~Rey, 
``String Theory on Thin Semiconductors,"
Progress of Theoretical Physics Supplement No. 177 (2009) pp. 128-142;
 arXiv:0911.5295 [hep-th].



\bibitem[{Sachdev \& Mueller(2008)}]{Sachdev:2008ba}
S.~Sachdev and M.~Mueller,
``Quantum Criticality and Black Holes,''
arXiv:0810.3005 [cond-mat.str-el].

\bibitem[{Sachdev(2010a)}]{Sachdev:2010ch}
  S.~Sachdev,
  ``Condensed matter and AdS/CFT,''
  arXiv:1002.2947 [hep-th].

\bibitem[{Sachdev(2010b)}]{Sachdev:2010um}
  S.~Sachdev,
  ``Holographic metals and the fractionalized Fermi liquid,''
  arXiv:1006.3794 [hep-th].



\bibitem[{Schafer \& Schwenzer(2004)}]{Schafer:2004zf}
  T.~Schafer and K.~Schwenzer,
  ``Non-Fermi liquid effects in QCD at high density,''
  Phys.\ Rev.\  D {\bf 70}, 054007 (2004)
  arXiv:hep-ph/0405053.
  %%CITATION = PHRVA,D70,054007;%%


\bibitem[{Shankar(1993)}]{Shankar:1993pf}
  R.~Shankar,
  %``Renormalization group approach to interacting fermions,''
  Rev.\ Mod.\ Phys.\  {\bf 66}, 129 (1994).
  %%CITATION = RMPHA,66,129;%%



\bibitem[{Silverstein(2004)}]{modulistabilization1}
  E.~Silverstein,
  ``TASI/PiTP/ISS lectures on moduli and microphysics,''
  arXiv:hep-th/0405068;
  
  \bibitem[{Son \& Starinets(2002)}]{Son:2002sd}
D.~T.~Son and A.~O.~Starinets,
``Minkowski-Space Correlators in AdS/CFT Correspondence: Recipe and   Applications,''
JHEP {\bf 0209} (2002) 042
[arXiv:hep-th/0205051].

\bibitem[{Spradlin \& Strominger(1999)}]{Spradlin:1999bn}
  M.~Spradlin and A.~Strominger, `Vacuum states for AdS(2) black holes,''
  JHEP {\bf 9911}, 021 (1999)
  [arXiv:hep-th/9904143].



\bibitem[{Varma \emph{et~al.}(1989)}]{varma}
  C.~M.~Varma, P.~B.~Littlewood, S.~Schmitt-Rink, E.~Abrahams and A.~E.~Ruckenstein,
  ``Phenomenology of the normal state of Cu-O high-temperature
  superconductors,''
  Phys.\ Rev.\ Lett.\  {\bf 63}, 1996 (1989).



\bibitem[{Vegh(2010)}]{Vegh:2010fc}
  D.~Vegh,
  ``Fermi arcs from holography,''
  arXiv:1007.0246 [hep-th].


\bibitem[{Witten(1998)}]{AdS/CFT2}
E.~Witten,
``Anti-de Sitter space, thermal phase transition, and confinement in  gauge theories,''
Adv.\ Theor.\ Math.\ Phys.\  {\bf 2} {\it ibid}. 505 (1998);


\bibitem[{Yamamoto \& Si(2010)}]{qimiao}
S.~J.~Yamamoto and Q.~Si, ``Global Phase Diagram of the Kondo Lattice: From Heavy Fermion Metals to Kondo Insulators,'' J. Low Temp. Phys. {\bf 161}, 233 (2010); arXiv:1006.4868.





\end{thebibliography}
\end{document}